\documentclass[fleqn,usenatbib]{mnras}
\usepackage{newtxtext,newtxmath}
\usepackage[T1]{fontenc}
\usepackage{ae,aecompl}

\usepackage{graphicx}	% Including figure files
\usepackage{amsmath}	% Advanced maths commands
\usepackage{amssymb}	% Extra maths symbols

\usepackage{xspace}
\usepackage{subfigure}
\usepackage{caption}
\usepackage{scrextend}
\usepackage{hyperref} 
\usepackage{pdflscape}

\newcommand{\esp}{\textit{ESPaDOnS}\xspace}
\newcommand{\narval}{\textit{NARVAL}\xspace}
\newcommand{\Rprime}{$R^{'}_{HK}$\xspace}
\newcommand{\caII}{Ca II H \& K\xspace}
\newcommand{\smw}{$S_{MW}$\xspace}

\title[Chromospheric emission of solar-type stars]{Chromospheric emission of solar-type stars with asteroseismic ages}

\author[R. S. Booth et al.]{R. S. Booth$^{1}$\thanks{rbooth03@qub.ac.uk}, K. Poppenhaeger$^{2,3,1}$, C. A. Watson$^{1}$, V. Silva Aguirre$^{4}$,
\newauthor
D. Stello$^{5,6,4}$ and H. Bruntt$^{7}$
\\
$^{1}$Astrophysics Research Centre, School of Mathematics \& Physics, Queen's University Belfast, University Road, Belfast BT7 1NN, UK\\
$^{2}$Leibniz Institute for Astrophysics Potsdam, An der Sternwarte 16, 14482 Potsdam, Germany\\
$^{3}$Institute for Physics and Astronomy, University of Potsdam, Karl-Liebknecht-Str. 24/25, 14476 Potsdam, Germany\\
$^{4}$Stellar Astrophysics Centre, Department of Physics and Astronomy, Aarhus University, Ny Munkegade 120, DK-8000 Aaarhus C, Denmark\\
$^{5}$School of Physics, University of New South Wales, Barker Street, Sydney, NSW 2052, Australia\\
$^{6}$Sydney Institute for Astronomy (SIfA), School of Physics, University of Sydney, NSW 2006, Australia\\
$^{7}$Aarhus Katedralskole, Skolegyde 1, 8000 Aarhus C, Denmark\\
}

\date{Accepted XXX. Received YYY; in original form ZZZ}
\pubyear{2019}

\begin{document}
\label{firstpage}
\pagerange{\pageref{firstpage}--\pageref{lastpage}}
\maketitle

% Abstract of the paper
\begin{abstract}
Stellar magnetic activity decays over the main-sequence life of cool stars due to the stellar spin-down driven by magnetic braking. The evolution of chromospheric emission is well-studied for younger stars, but difficulties in determining the ages of older cool stars on the main sequence have complicated such studies for older stars in the past. Here we report on chromospheric Ca~II H and K line measurements for 26 main-sequence cool stars with asteroseismic ages older than a gigayear and spectral types F and G. We find that for the G stars and the cooler F-type stars which still have convective envelopes the magnetic activity continues to decrease at stellar ages above one gigayear. Our magnetic activity measurements do not show evidence for a stalling of the magnetic braking mechanism, which has been reported for stellar rotation versus age for G and F type stars. We also find that the measured \Rprime indicator value for the cool F stars in our sample is lower than predicted by common age-activity relations that are mainly calibrated on data from young stellar clusters. We conclude that, within individual spectral type bins, chromospheric magnetic activity correlates well with stellar age even for old stars.
\end{abstract}

% Select between one and six entries from the list of approved keywords.
% Don't make up new ones.

\section{Introduction}

All cool stars exhibit features of magnetic activity -- a collective term for phenomena related to the stellar magnetic field, such as flares, star spots, and the existence of a chromosphere and corona. The stellar magnetic field is driven by the stellar rotation, which transforms rotational energy into magnetic fields through a dynamo process \citep{1955ApJ...122..293P}. The overall magnetic activity level of cool stars decreases over time; the observed age-activity relationship stems from the rotational evolution of cool stars during their main-sequence lifetime. These stars spin-down over time due to magnetic braking, which removes angular momentum via a magnetised stellar wind \citep{1962AnAp...25...18S}. The decrease in rotational velocity via magnetic braking consequently causes a reduction in the magnetic field generated by the stellar dynamo, which can be observed through magnetic activity proxies, one of which is emission in spectral lines produced in the stellar chromosphere.

The use of chromospheric emission from the \caII lines as a proxy for magnetic activity in stars has a long history, dating back to the HK project at the Mount Wilson Observatory \citep{1968ApJ...153..221W} where it was used to search for stellar analogues of the solar cycle; for more recent studies, see for example \citet{2007AJ....133..862H, 2009AJ....138..312H}. It has also been widely used to study the age-activity relationship \citep{1972ApJ...171..565S,1991ApJ...375..722S,2008ApJ...687.1264M,2013A&A...551L...8P}. \citet{1972ApJ...171..565S} was the first to plot calcium emission as a function of age ($t$) and found a relationship that was proportional to $t^{-\frac{1}{2}}$. Since then, efforts have been made to calibrate the age-activity relationship across spectral types using the \Rprime indicator, introduced by \citet{1984ApJ...279..763N}. The \Rprime indicator is calculated from the S-index, which is a measurement of the \caII emission strength normalized to the continuum. A colour correction is applied to the S-index and then the underlying photospheric contribution is removed to obtain the \Rprime indicator.

There have been several studies of the age-activity relationship \citep{1991ApJ...375..722S,1999A&A...348..897L,2008ApJ...687.1264M}, however, these lack stars older than a gigayear with reliable and accurate ages. In recent years, there have been efforts to improve the age-activity relationship beyond a gigayear. \citet{2013A&A...551L...8P} investigated the relationship using field stars and found a L-shaped distribution that indicates that the decay of chromospheric activity stops relatively early in the main sequence ($\approx 1.5$ Gyr). However, this study used isochrone ages for their sample with typical errors around 2 Gyr or up to 30\% of the stellar age. It is also known that metallicity has an effect on the value of \Rprime indicator calculated \citep{1998MNRAS.298..332R}. Another recent study by \citet{2016A&A...594L...3L} took into account mass and metallicity into the age-activity relationship and found good agreement with asteroseismic ages.

In order to study the age-activity relationship beyond a gigayear, it is crucial to obtain the best estimate for the ages of the stars. While clusters have been used previously to investigate the relationship \citep{2008ApJ...687.1264M,2013A&A...551L...8P}, there are very few clusters older than a gigayear that are easy to study with high-resolution spectra for \caII measurements. Such clusters tend to be more distant making analysis of magnetic activity proxies more difficult. Historically, using field stars to calibrate the age-activity relationship has been difficult because most age-dating methods work best for younger stars ($< 1$ Gyr). However, with the advancement of space telescopes such as \textit{CoRoT} and \textit{Kepler}, it has been possible to determine fundamental parameters such as the ages of stars through asteroseismology \citep{2013ARA&A..51..353C, 2012ApJ...749..152M, 2012ApJ...748L..10M, 2014ApJS..210....1C, 2015MNRAS.452.2127S}. This technique provides information about the interior of stars through observations of stellar oscillations and has proved to be the most accurate age-dating method for old field stars. Asteroseismology has opened up the possibility of stellar age investigations for stars older than a gigayear with recent studies using the technique to study rotation \citep{2016Natur.529..181V} and coronal X-ray activity \citep{2017MNRAS.471.1012B}.

In this work we use asteroseismic ages along with Ca~II \Rprime values to investigate the age-activity relationship for stars older than a gigayear. In Section \ref{observations} we present the data used to investigate the relationship, followed by Section \ref{method} that details the analysis performed on the data. Section \ref{results} presents the results from the study and Section \ref{discussion} provide the discussion for these results. Finally, Section \ref{conclusions} summarises the conclusions from this work.

\section{Observations}
\label{observations}

\subsection{Sample selection}
\label{sample_selection}

We selected our sample from stars studied by \citet{2012MNRAS.423..122B}, who performed a detailed spectral study aided by asteroseismic data for 93 solar-type stars observed with Kepler. To study the magnetic activity of old stars, we restricted our sample to stars that have an outer convective envelope, are on the main sequence, and have an asteroseismically determined age.

To select stars with an outer convective envelope, we required the stellar effective temperature to be below 6200~K, equivalent to spectral type $\sim$F7V. Stars above this temperature lie above the Kraft break \citep{1967ApJ...150..551K} and are observed to be rapidly rotating (e.g. \citealt{1997PASP..109..759W,2012A&A...537A.120Z}). At this point, the outer convective zone becomes extremely thin and is unable to generate the magnetic winds needed for angular momentum loss on the main sequence. Therefore, stars hotter than 6200~K do not follow the age-activity relationship. However, there is some observational evidence that hotter stars can occasionally display magnetic activity features as well; for example, weak X-ray emission was detected from the A7V star Altair \citep{2009A&A...497..511R}. We have therefore performed our data analysis on all main-sequence stars in the sample spanning the full effective temperature range of $\sim$5000--6900\,K, but display the hotter stars with effective temperatures between 6200 and 6900\,K separately in our tables and figures.

To ensure that the stars in the sample are on the main sequence, we compared their surface gravity \citep{2012MNRAS.423..122B} from asteroseismology to the relation between $B-V$ colour and surface gravity for main-sequence stars, as given by \cite{CBO9781316036570A207}, and excluded stars from our sample that differed by more than 0.2~dex . Stars that have significantly evolved off the main sequence are expected to have a different rotational and magnetic evolution, which is why we focused only on main-sequence stars in our analysis. 

Stellar ages derived from asteroseismology were collected from the literature. In particular, we used ages from \citet{2017ApJ...835..173S} for the majority of our sample, where stellar ages were obtained by modelling the individual oscillation frequencies in the spectrum observed by the \textit{Kepler} satellite. Some ages were also taken from \citet{2014ApJS..210....1C}, where stellar properties were estimated using two global asteroseismic parameters and complementary photometric and spectroscopic data. The full sample of stars are presented in Appendix \ref{calcium_results}.

\subsection{Spectra}
The spectra used in this study stem from \citet{2012MNRAS.423..122B}. The spectra were obtained with the \esp spectrograph at the 3.6-m Canada-France-Hawaii Telescope (CFHT; \citealt{2006ASPC..358..362D}) and the \narval spectrograph \citep{2003EAS.....9..105A} mounted on the 2-m Bernard Lyot Telescope at the Pic du Midi Observatory. These spectra cover a spectral range of ${\thicksim} 3700$ to ${\thicksim} 10480$ \AA \space and have a nominal resolution of $R = 80 000$. We used the reduced spectra by \citet{2012MNRAS.423..122B}, which we received through personal communication. The reduction process is described in detail in \citet{2012MNRAS.423..122B}.

Additional archival observations were used to investigate the level of long-term variability of the \Rprime indicator due to potential magnetic activity cycles (see Section \ref{two_channel_analysis} and \ref{stellar_variability} for details). Pipeline-reduced spectra were obtained directly from the \esp archive\footnote{\url{http://www.cadc-ccda.hia-iha.nrc-cnrc.gc.ca/en/}}. The details of these additional observations are listed in Table \ref{additional_obs_table}.

\begin{table}
\centering
\caption{Details of observations taken from \esp archive to sample potential magnetic activity cycles present.}
\label{additional_obs_table}
\begin{tabular}{lclc}
\hline
Star Name    & Year & Product ID & Integration Time / s \\
\hline
KIC 6106415  & 2014 & 1740569i   & 60                   \\
KIC 8694723  & 2014 & 1733346i   & 255                  \\
KIC 9139151  & 2014 & 1733349i   & 280                  \\
KIC 9955598  & 2013 & 1630041i   & 1700                 \\
KIC 10454113 & 2014 & 1733354i   & 165                  \\
KIC 10454113 & 2016 & 2005115i   & 840                  \\
KIC 10963065 & 2013 & 1628074i   & 1625                 \\
\hline
\end{tabular}
\end{table}

\section{Data analysis}
\label{method}

Here we describe how the collected optical spectra were analysed to extract the chromospheric emission in the Ca~II H \& K lines, how this was calibrated with respect to the Mount Wilson S index, and how the final \Rprime activity indicator was calculated.

\subsection{Doppler correction}
To make the extraction of the Ca~II H \& K line emission easier, we first corrected for any Doppler shift that was present in the individual stellar spectra. A master spectrum for each spectrograph was chosen based on the signal to noise ratio (SNR) in the 3850 to 4050 \AA \space wavelength region. The master spectra were KIC 9226926 and KIC 3733735 for the \narval and \esp spectra, respectively. All other spectra were then corrected by using a cross correlation function with respect to the master spectrum. An additional manual wavelength correction was applied to the spectra in order to compensate for any Doppler shift that was present in the master spectrum. This was -0.2 \AA \space for the \esp data; the \narval data did not require this manual correction.

\subsection{Normalization of continuum}
\label{spectra_normalization}
\begin{figure}
	\includegraphics[scale=0.25]{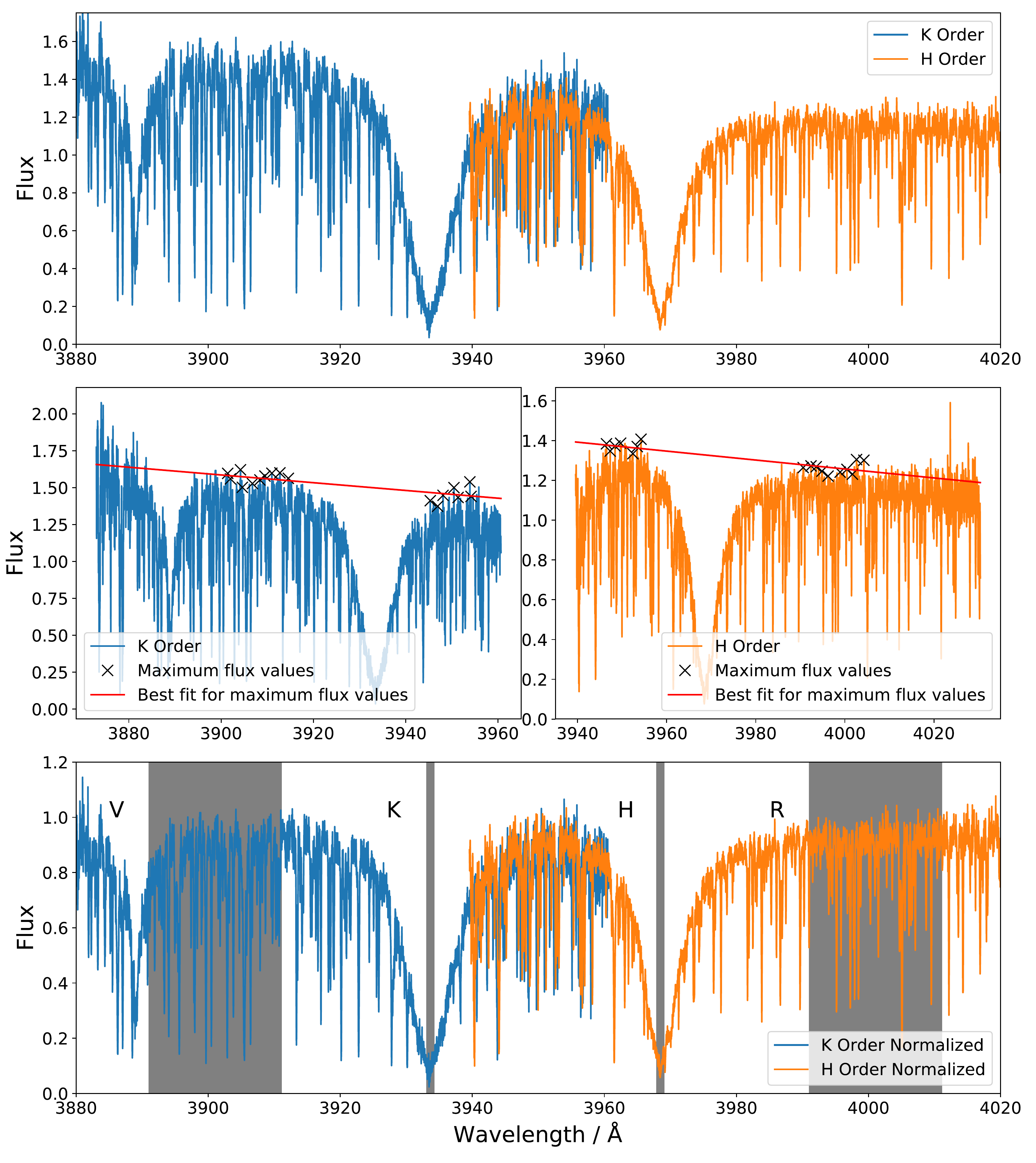}
	\caption{\textbf{Top panel:} Spectrum of KIC 9139163 showing the two adjacent spectral orders that contain the \caII lines. While the overlapping region of the orders match, there is a clear discontinuity in the continuum flux level.
    \textbf{Middle panel:} Plots of the two comparison regions in the K order (left) and H order (right). The markers denote the maximum flux values in each bin of width 1.5 \AA. The red line denotes the best-fitting linear relationship that is used to normalize the spectrum.
    \textbf{Bottom panel:} Spectrum of KIC 9139163 with normalization relationship applied to both orders. The discontinuity in the continuum flux level is no longer present. The four channels used to calculate the S index are also shown in grey.}
	\centering
	\label{normalisation_method}
\end{figure}

Before the \Rprime indicator could be calculated, the spectra were required to be normalized as there was a discontinuity in the continuum level at the \caII wavelengths. This is shown in the top panel of Figure \ref{normalisation_method}; the flux in the overlapping region of the two orders match reasonably well, but the continuum flux level is higher in the K order than in the H order. This would have an affect on the chromospheric emission measurement as the method used (see Section \ref{s_method}) requires the flux measured in the core of the \caII line to be normalized by the continuum flux in two reference channels.

This was corrected by considering two comparison regions in each spectral order; from 3900 - 3915 \AA \space and 3945 - 3955 \AA \space in the shorter-wavelength order of the spectrum (containing the Ca~II K line) and from 3945 - 3955 \AA \space and 3990 - 4005 \AA \space in the longer-wavelength order of the spectrum (containing the Ca~II H line) as shown in the middle panel of Figure \ref{normalisation_method}. We estimated the local continuum level in those regions by splitting the spectrum into bins of 1.5 \AA \space width and measuring the local maximum in each of the bins (see middle panel of Figure \ref{normalisation_method}). These data points were used to find the best-fitting linear relationship as indicated by the red line in the middle panel of Figure \ref{normalisation_method}. The spectra were then normalized according to this best fitting relationship so that the typical continuum flux value in each order was approximately one as shown in the bottom panel of Figure \ref{normalisation_method}.

\subsection{Calculation of the S index}
\label{s_method}
The S index is an activity indicator that uses four channels to quantify the chromospheric emission in the core of the \caII lines. Two channels are 1.09 \AA \space wide and centred on 3968.47 and 3933.664 \AA \space for the \caII cores, respectively. The other two channels are 20 \AA \space wide and are centred on 3901.07 and 4001.07 \AA; named the V and R channels, respectively. These channels are shown in the bottom panel of Figure \ref{normalisation_method} and follow the channels defined by \citet{2011arXiv1107.5325L} with the exception that we do not include the triangular-shaped profile to the H and K channels.

The S index is calculated using Equation \ref{sindex} where $N_{x}$ is the total flux in the relevant channel. The error associated with $N_{x}$ is calculated using Equation \eqref{sindex_error} where $x_{n}$ is the error on the individual flux data point and $i$ is the total number of data points in the wavelength channel. These errors associated with $N_{x}$ were then propagated in order to calculate the error associated with the S index.

\begin{equation}
S = \frac{N_{H} + N_{K}}{N_{R} + N_{V}}
\label{sindex}
\end{equation}

\begin{equation}
\sigma_{N_{x}} = \sqrt{\sum_{n=1}^{i} x_{n}^{2}}
\label{sindex_error}
\end{equation}

\subsection{Calibration to \texorpdfstring{\smw}{Smw} and calculating the \texorpdfstring{\Rprime}{Rprime} indicator}
\begin{figure}
	\centering
	\subfigure[\esp calibrators]{\includegraphics[scale=0.55]{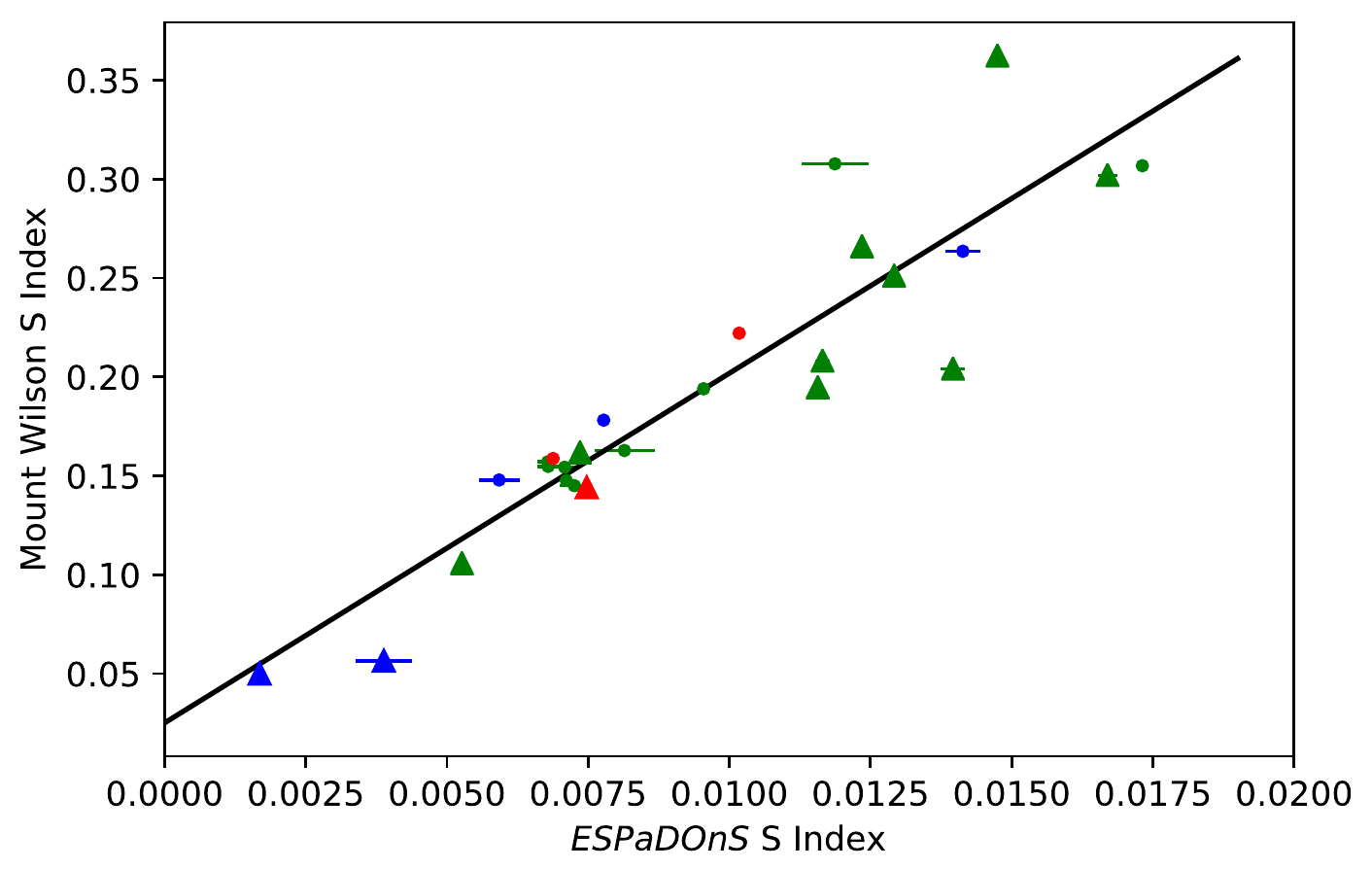}}\\
    
    \subfigure[\narval calibrators]{\includegraphics[scale=0.55]{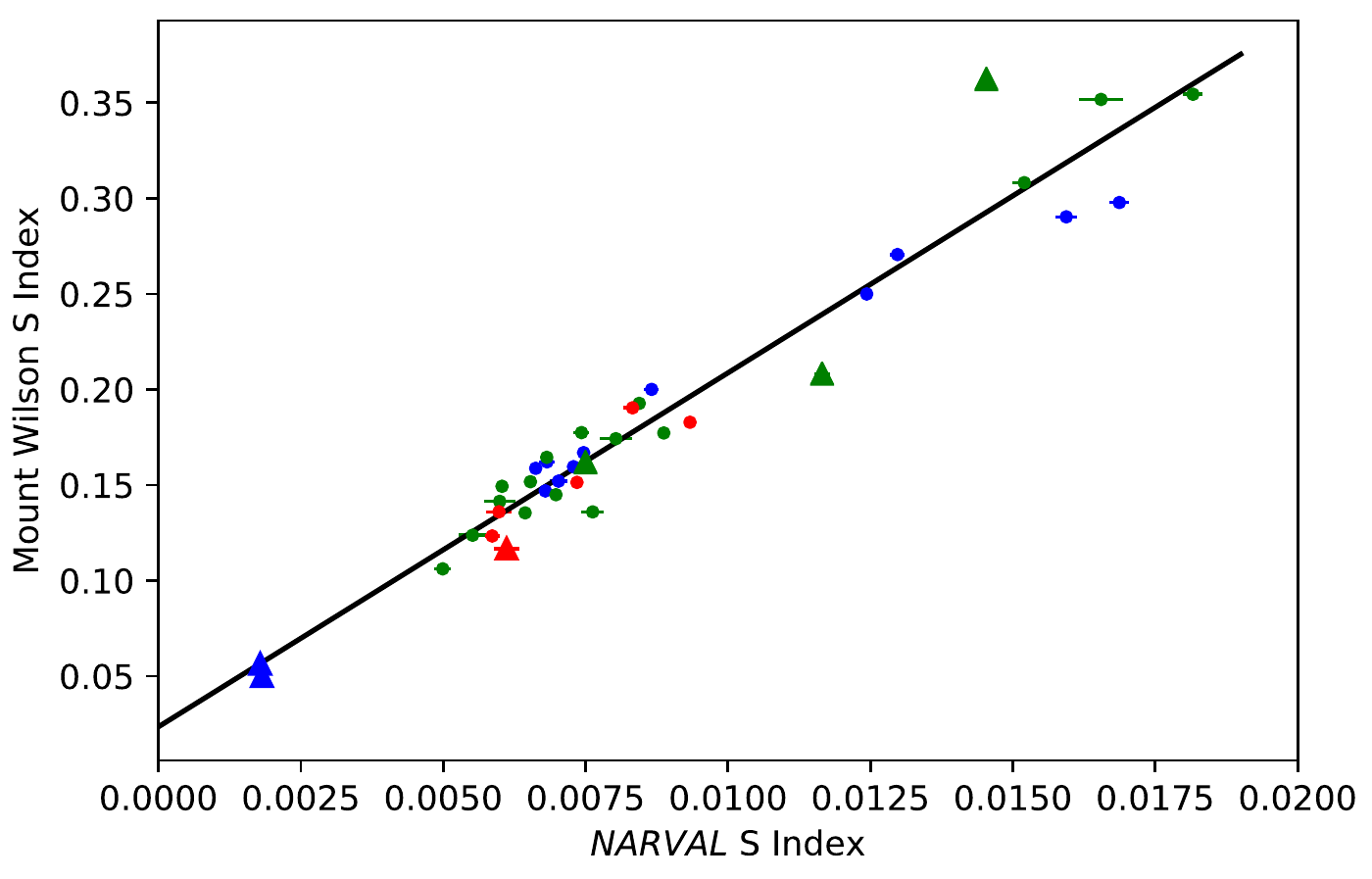}}
    
    \caption{Plots showing the average \smw \citep{1991ApJS...76..383D} against the S index calculated from the \esp/\narval spectrographs for the calibrator stars. Triangle markers indicate giant stars and circle markers represent main sequence stars. Blue indicates F-type stars, green indicates G-type stars, and red indicates K-type stars. The black line indicates the best-fitting linear relationship between the two parameters.}
	\label{nar_and_esp_calibrator_plot}
\end{figure}

The method that is used to calculate the \Rprime indicator \citep{1984ApJ...279..763N} requires the Mount Wilson S index (\smw), which differs slightly from the S index that is calculated for our sample of stars. The S index calculated in any spectrum (regardless of the exact method used) is dependent on the spectral resolution and the throughput of the spectrograph used. Therefore it was necessary to determine a relationship between the S index calculated from the spectrographs used in this work (\narval and \esp) and \smw. This was achieved by searching for stars from \citet{1991ApJS...76..383D} in the \narval and \esp archives thus providing a calibrator sample. \citet{1991ApJS...76..383D} summarizes measurements made of \caII emission at the Mount Wilson Observatory from 1966 - 1983, thus providing a large sample of stars with measured \smw values. We restricted the calibrator sample to have the same range of spectral types as used in our analysis, namely spectral type later than $\sim$F7. The S index was calculated for these calibrator stars as described in Section \ref{s_method} which included the Doppler correction and normalization of the continuum. The master spectra for the Doppler correction of the calibrator star sample were HD 89449 and HD 126660 for the \esp and \narval data, respectively. The manual wavelength shift was -0.14 and +0.1 \AA \space for the \esp and \narval data, respectively. The full list of stars and values used to calibrate the two relationships are shown in Appendix \ref{list_calibrator_stars}. Note that some stars in the calibrator sample are classed as giant stars (and are denoted by triangle markers in Figure \ref{nar_and_esp_calibrator_plot}). While these stars are not relevant for the age-activity relationship, they provide additional data on the relationship between the two indices particularly where there are a limited number of main sequence stars as seen in the \esp relationship.

The S index calculated in each spectrograph is plotted against the mean \smw from \citet{1991ApJS...76..383D} (shown in Figure \ref{nar_and_esp_calibrator_plot}). Since we are considering weakly inactive stars older than a gigayear, we restricted the calibrator star sample to S index values less than 0.2 and \smw values less than 0.5. A linear least-squares regression was applied to the data to find the best-fitting relationship between the two indices, which are shown in Equations \eqref{esp_calibrator_eq} and \eqref{nar_calibrator_eq}. These equations were used to calculate \smw for our sample of stars with ages from asteroseismology.

\begin{equation}
S_{MW} = 17.67 \cdot S_{\mathrm{ESPaDOnS}}+ 0.025
\label{esp_calibrator_eq}
\end{equation}

\begin{equation}
S_{MW} = 18.53 \cdot S_{\mathrm{NARVAL}} + 0.023
\label{nar_calibrator_eq}
\end{equation} 

Using the \smw values calculated from the calibration relationships, the \Rprime indicator was calculated using the original method by \citet{1984ApJ...279..763N}. The method from \citet{1984ApJ...279..763N} requires the stellar $(B-V)$ colour. For our sample of stars, these values were obtain from SIMBAD \citep{2000A&AS..143....9W}. The effect of reddening on the value of \Rprime indicator was investigated by comparing stellar positions and \textit{Gaia} DR2 distances to reddening maps from \citet{2019A&A...625A.135L}\footnote{Available at \url{http://stilism.obspm.fr}}. The \Rprime indicator was calculated using both the observed $(B-V)$ from SIMBAD and the intrinsic $(B-V)_{0}$ that takes reddening into account. The average difference between the $\log$\Rprime indicator was $\sim 0.003$~dex. Therefore, reddening has a negligible effect on the value of the \Rprime indicator calculated and the observed $(B-V)$ from SIMBAD is used in this work.

%\hspace{-0.2cm}

\subsection{Two channel analysis}
\label{two_channel_analysis}

Additional observations were searched for in the \esp archive in order to investigate the effect of potential magnetic activity cycles on the value of the \Rprime indicator (see Section \ref{stellar_variability}). In the \esp archive six stars (with $T_{eff} < 6200$ K) were found to have additional observations, the details of these observations are shown in Table \ref{additional_obs_table}. Note that no additional observations were found in the \narval archive. However, a significant number of spectra taken from the \esp archive for this part of the analysis contained strong scatter in the spectral order where the Ca II K line is located, whereas the scatter was much lower in the spectral order containing the Ca II H line. Therefore, a modified S index and \Rprime indicator using only the H and R channels was calculated. Specifically, the modified S index was calculated using Equation \ref{Eq:modified_S} where $N_{x}$ is the total flux in the relevant channel.

\begin{equation}
    S_{mod} = \frac{N_{H}}{N_{R}}
    \label{Eq:modified_S}
\end{equation}

Since the Ca II H typically shows slightly less of a flux excess in the line centre than the K line, a new \smw calibration was performed for the modified S index using the calibrator stars from \citet{1991ApJS...76..383D}. This modified S index was calculated for the calibrator stars and plotted against the mean \smw from \citet{1991ApJS...76..383D}. Similarly to the four channel analysis, the sample was restricted to S index values less than 0.2 and \smw values less than 0.5. The best-fitting relationships for the modified S index and average \smw value for each spectrograph are shown in Equation \ref{Eq:esp_calibrator_eq_mod} and \ref{Eq:nar_calibrator_eq_mod}.

\begin{equation}
S_{MW} = 21.93 \cdot S_{\mathrm{ESPaDOnS (mod)}}+ 0.016
\label{Eq:esp_calibrator_eq_mod}
\end{equation}

\begin{equation}
S_{MW} = 21.79 \cdot S_{\mathrm{NARVAL (mod)}} + 0.014
\label{Eq:nar_calibrator_eq_mod}
\end{equation}

Once the Mount Wilson S value has been found, the conversion to the \Rprime indicator is identical to the four channel method. This two channel method is used when data from the \esp archive is needed as these spectra have strong scatter in the spectral order containing the Ca II K line. Therefore, the two channel method is used when considering the potential stellar variability (see section \ref{stellar_variability}). The results of these additional observations are shown in Appendix \ref{multiple_obs_tables_appendix}.

\section{Results}
\label{results}

\begin{figure*}
	\includegraphics[width=0.99\textwidth]{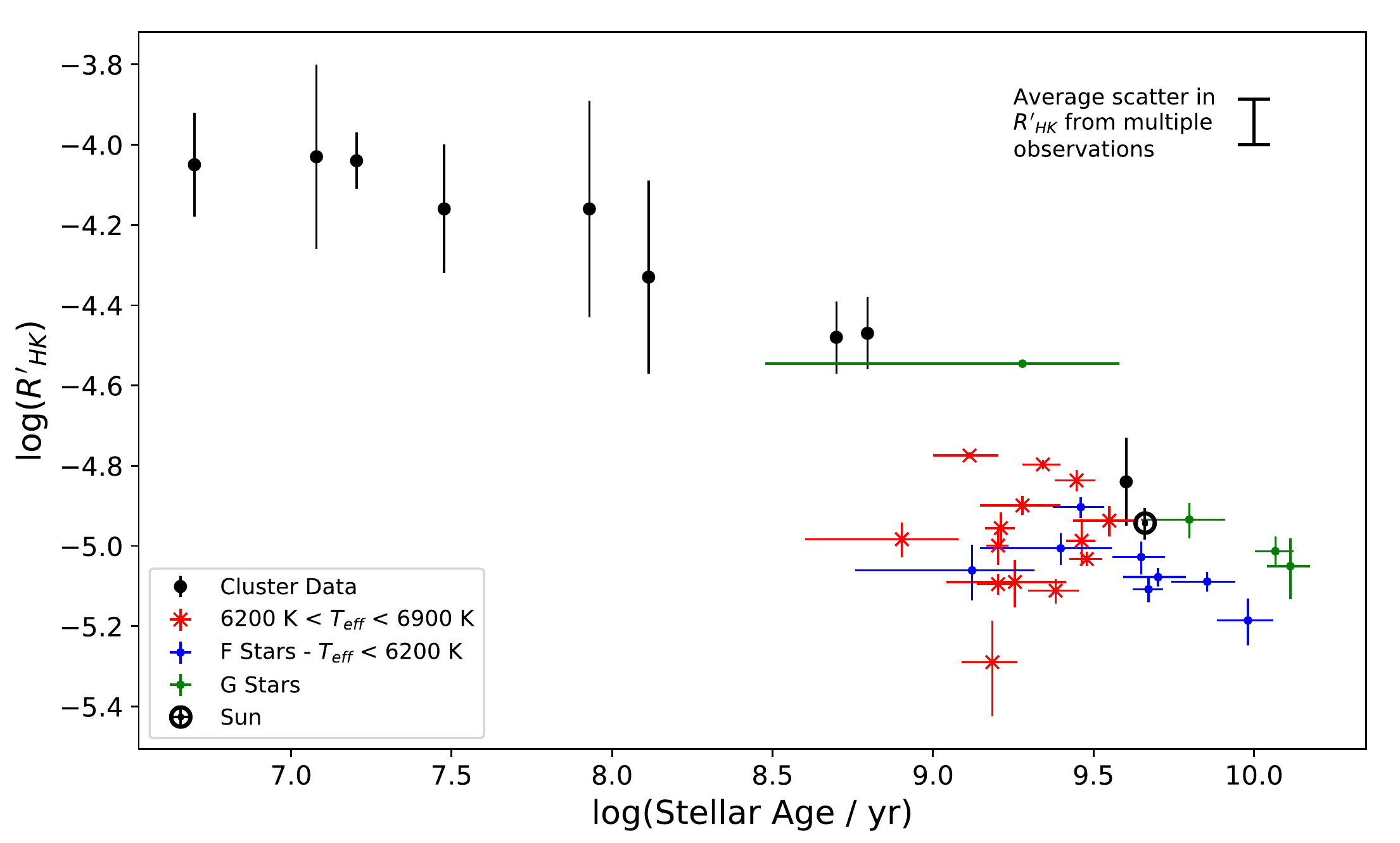}
	\caption{Plot showing data analysed in this work alongside cluster data from \citet{2008ApJ...687.1264M} shown in black. Early F-ype stars (with $T_{eff} > 6200$ K) are shown as red markers, later F-type stars (with $T_{eff} < 6200$ K) are denoted by blue markers and G-type stars are denoted by green markers. The average solar value over several cycles is also shown \citep{2017ApJ...835...25E}.}
	\centering
	\label{calcium_emission_plot}
\end{figure*}

\subsection{Comparison to cluster data}
\label{comparison_to_cluster}
We measured the chromospheric emission in the Ca~II H\&K lines in the spectra of 26 stars with asteroseismic ages. Our sample contains fourteen early F-type stars with effective temperatures between 6200 and 6900 K, eight late F-type stars and four G-type stars. As described in Section~\ref{sample_selection}, we used the measured surface gravities \citep{2012MNRAS.423..122B} to select stars that are on or near to the main sequence. In Figure \ref{calcium_emission_plot} we display the \Rprime activity indicator as a function of stellar age. We colour-coded different groups of stars in the plot; specifically, we display G-type stars in green, late F stars in blue, and earlier F-type stars in red.

The full details for each star are given in Appendix \ref{calcium_results}. Three stars (KIC 5774694, KIC 9139163 and KIC 10454113) had observations from both \narval and \esp, the values found for the \Rprime indicator from each spectrograph are consistent with one another. Therefore, for simplicity, only the \narval data for these stars are shown in Figure \ref{calcium_emission_plot} as these have smaller associated errors in the activity measurement.

To set the data for our old stars into context, we also show the typical spread of chromospheric activity in young stellar clusters and the old cluster M67 in Figure \ref{calcium_emission_plot}. The cluster data is taken from Table 6 of \citet{2008ApJ...687.1264M} with errors representative of the 68\% confidence level; \citet{2008ApJ...687.1264M} collected stellar \Rprime indicators from the available literature in a colour range of $(B-V) = 0.46$ to $0.89$, which corresponds to effective temperatures of 5000 -- 6500\,K and matched closely our sample selection for G, and late F dwarfs. For further comparison,  we also show the mean \Rprime value for the Sun over several cycles \citep{2017ApJ...835...25E}; the age of the Sun is taken to be $4.57 \pm 0.02$ Gyr \citep{1995RvMP...67..781B}.

The early F-type stars (shown in red in Figure \ref{calcium_emission_plot}), do not display a strong correlation between activity and age with a calculated Pearson correlation coefficient of 0.098. This is to be expected since these stars lie above the Kraft break and therefore do not generate the magnetic winds needed for angular momentum loss on the main sequence. This confirms the effective temperature limit for the age-activity relationship. However, these stars still display signs of chromospheric activity with a range of \Rprime values primarily between $-5.1< \mathrm{log}(R^{'}_{HK})< -4.8$. Therefore, while these stars cannot give insight into the age-activity relationship, they show that stars above the Kraft break are still magnetically active.

Concerning the cooler part of the sample (i.e. the late F-type and G-type stars), Figure \ref{calcium_emission_plot} shows that for each of these spectral types, there is a strong correlation between activity and age as expected. This is quantified by calculating the Pearson correlation coefficient; for the late F-type stars this value is $-0.582$ and for the G-type stars this value is $-0.986$. Considering the small sample considered in this work, the true correlation coefficients may be different than these values, particularly for the G-type stars since the sample only comprises four stars. One star in our sample, the G dwarf KIC 5774694, has an unusually large \Rprime value; it is also a star with a large uncertainty in its asteroseismic age determination. Unfortunately, we only have a single epoch of an activity measurement for this star so that we cannot test if a stellar flare temporarily increased the activity of this star. Therefore, we consider this star as a potential outlier with respect to age or activity.

The cluster M67 has an age of $\sim 4$\,Gyr \citep{1992AJ....103..151D,2004PASP..116..997V,2010A&A...513A..50B} 
and the chromospheric activity of $\sim$70 of its member stars has been reported in the literature \citep{2006ApJ...651..444G,2008ApJ...687.1264M}. The cluster is well-studied and measurements of the stellar rotation periods make it a benchmark for the spin-down of cool stars \citep{2016ApJ...823...16B}. It is therefore a useful comparison target to our old-star sample, and we display the mean and variance of its stellar chromospheric activity, as reported by \citealt{2008ApJ...687.1264M}, in our Figure~\ref{calcium_emission_plot}.

At a first glance the chromospheric activity of the M67 stars appears higher than what we see in our sample of cool stars for similar ages. However, it is important to note that \citealt{2008ApJ...687.1264M} calculated the \Rprime values from the measurements performed by \citealt{2006ApJ...651..444G}, who studied only G-type stars in their work. So the appropriate comparison group is the old G stars in our sample, i.e.\ the green data points in Figure~\ref{calcium_emission_plot}. There the discrepancy disappears: the G stars in our sample, while being somewhat older than M67, follow the trend of M67 and younger clusters. However, recent work from \citet{2017AJ....153..275C} discusses the effect of the interstellar medium (ISM) on the \caII lines in relation to the M67 cluster. Therefore, it is possible that the true \Rprime indicator values for the M67 stars are slightly higher than the values reported in \citet{2006ApJ...651..444G} when corrected for the ISM. However, \Rprime corrections for individual stars were not given in their work, so we do not attempt to perform a numerical adjustment for the M67 stars in our work.

The F stars however have significantly lower activity than the G dwarfs in M67. This seems reasonable, since F stars at that age are closer to the end of their main-sequence lifetime than G stars of the same age. Nevertheless, in the light of recent work which reported a stalling in the spin-down for old G and F main-sequence stars \citep{2016Natur.529..181V}, we decided to test if our very low activity measurement might be caused be differences in the data analysis, compared to younger clusters. We therefore conducted an additional analysis to see if data reduction issues could be the reason for this discrepancy. To this aim, we searched for spectra of member stars of the three oldest clusters from \citealt{2008ApJ...687.1264M}, i.e.\ M67, the Ursa Major (UMa) moving group, and the Hyades, in the \esp and \narval archives. Unfortunately we did not find any archival observations of M67 member stars, but we found spectra of four UMa member stars and two Hyades member stars. We analysed those spectra with the same techniques as employed for our sample stars and compared the results to the listed \Rprime values in \citealt{2008ApJ...687.1264M}. We found some scatter in the values, which is not surprising given the intrinsic stellar variability of cool stars, but no systematic offset between our values and the literature values. We deduce that the very low activity levels for F stars in our sample are real, and discuss this in Section~\ref{spindown}.

\begin{figure*}
	\includegraphics[width=0.99\textwidth]{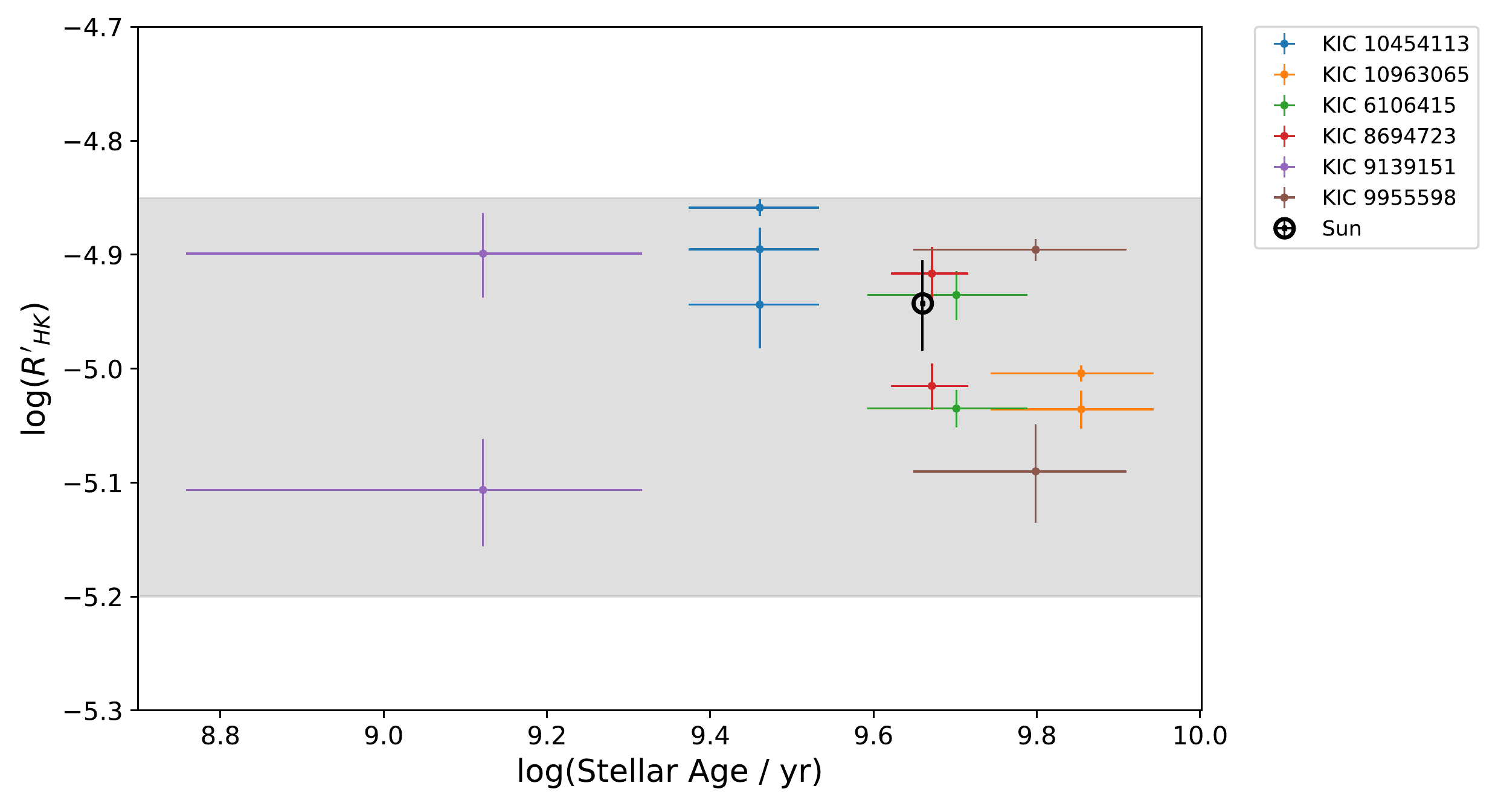}
	\caption{Plot showing the \Rprime indicator values calculated using only the H and R channels (see Section \ref{stellar_variability} for further details) for multiple observations of stars in our sample ($T_{eff} < 6500$ K). The grey shaded region indicates the range of the \Rprime indicator values that our sample have when the \Rprime indicator is calculated using only the H and R channels (see Appendix \ref{sample_2_channels} for plot). The average solar value over several cycles \citep{2017ApJ...835...25E} is also shown for reference.}
	\centering
	\label{multiple_obs_plot}
\end{figure*}

\subsection{Variability of individual stars in the sample}
\label{stellar_variability}
It is known that cool stars may have magnetic activity cycles similar to the solar eleven year activity cycle \citep{1978ApJ...226..379W,1995ApJ...438..269B}. We therefore investigated the effect of intrinsic stellar variability by sample stars being in different stages of a possible activity cycle. As described in Section \ref{two_channel_analysis}, additional observations were searched for in the \esp and \narval archives and six cool stars were observed at additional epochs.

The results of these additional observations along with the original observation for the relevant stars are shown as a function of their age in Figure \ref{multiple_obs_plot} (note that the \Rprime indicator values were recalculated for the original observation using only the H and R channels for consistency). Solar values are also shown for reference \citep{2017ApJ...835...25E}. The grey shaded region on the plot represents the range of $\log R^{'}_{\mathrm{HK,\,mod}}$ values of our whole sample over all investigated asteroseismic ages (see also Appendix \ref{sample_2_channels} for modified activity versus age plot). 

The plot shows that the typical intrinsic variability of individual stars with two and in one case three observations is in the range of 0.01 to 0.21 dex, with an average activity difference of 0.11 dex per individual star. The typical time separation of epochs is between 3 and 6 years, with the original spectra used by \citet{2012MNRAS.423..122B} having been collected in 2010. It is interesting to note that the F-type star KIC 6106415 that seemed to have a fairly low activity ($\log R^{'}_{\mathrm{HK}} = -5.035$) for its age ($1.07$~Gyr), at a later epoch has an activity indicator value of $\log R^{'}_{\mathrm{HK}} = -4.935$. This value for the \Rprime indicator at a later epoch brings it in line with the values observed for F-type stars of a similar age in our sample.

If the spread of the full late-F to G star sample was only due to intrinsic stellar variability, then the average pair-wise difference of \Rprime values in the sample should be close to the 0.11 dex variation which is typically displayed by individual stars with multiple observations in our sample. We find that the pairwise average activity difference, excluding the high-activity outlier star KIC~5774694 discussed in Section~\ref{comparison_to_cluster}, is 0.097 dex. This implies that the scatter in activity is slightly less than expected from intrinsic stellar variability.

\subsection{Planet-hosting stars in the sample}
From theoretical considerations \citep{2000ApJ...533L.151C}, stars with exoplanets in close orbits may experience enhanced stellar activity through star-planet interaction. This effect has been observed for some systems with high-mass exoplanets in very close orbits (see e.g. \citealt{2014A&A...565L...1P,2015ApJ...805...52P}). Two of the stars in the sample have confirmed exoplanets, KIC 9955598 (Kepler-409b) and KIC 10963065 (Kepler-408b). However, Kepler-408b is a small planet with less than five Earth masses in a ca.\ 2.5-day orbit, and Kepler-409b has a wider orbit with ca.\ 69-day period \citep{2014ApJS..210...20M}. Therefore it is unlikely that those two planets have a significant influence on their host stars' magnetic activity, and star-planet interactions are not expected to play a role in our investigation.

\section{Discussion}
\label{discussion}

\subsection{Comparison to existing age--activity relationships}
\label{previous_relationships}
There have been numerous studies that have calibrated the relationship between the \Rprime indicator and stellar age. Here, we will compare these previous relationships to the sample of cool stars analysed in this work.

\begin{figure}
	\includegraphics[scale=0.35]{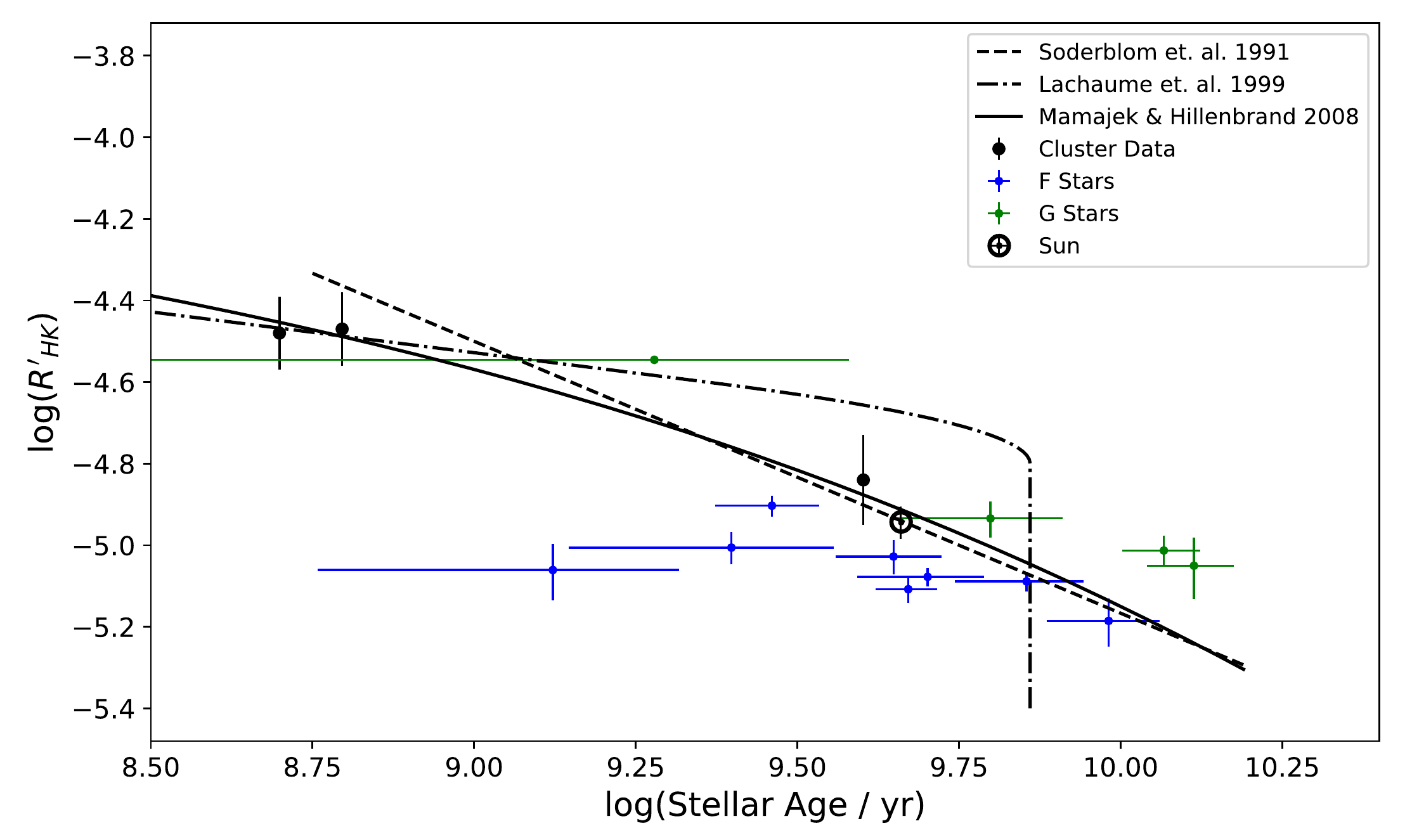}
	\caption{Plot showing three previously published calibrated relationships between the \Rprime indicator and age \citep{1991ApJ...375..722S,1999A&A...348..897L,2008ApJ...687.1264M} alongside the cluster data from \citet{2008ApJ...687.1264M} and the old sample of stars from this work. The average solar value over several cycles is also shown \citep{2017ApJ...835...25E}.}
	\centering
	\label{3_relationship_plot}
\end{figure}

The relationships that we consider are taken from \citet{1991ApJ...375..722S}, \citet{1999A&A...348..897L} and \citet{2008ApJ...687.1264M}; we show these three relationships alongside the sample of old cool stars analysed in this work in Figure \ref{3_relationship_plot}. The first relationship from \citet{1999A&A...348..897L} is consistent with the two youngest clusters shown in Figure \ref{3_relationship_plot}. However, this relationship suggests that the \Rprime indicator does not decay as rapidly as the other two relationships. It is worth noting that they have restricted their sample to $B-V > 0.6$ which may explain some of the disagreement with the other two relationships. In addition to this, some of their ages are not independent as they have been derived from rotation. The relationship from \citet{1991ApJ...375..722S} shown is the simple power law that they fitted to their data. We see that this relationship has reasonable agreement with the \citet{2008ApJ...687.1264M} relationship from $\approx$ 1 Gyr. However, it should also be noted that \citet{1991ApJ...375..722S} made an additional fit that included a constraint for constant star formation. This improved the relationship for younger ages but does not affect stars older than the Sun. The third relationship shown in Figure \ref{3_relationship_plot} is from \citet{2008ApJ...687.1264M}. As with the relationship from \citet{1991ApJ...375..722S}, there is good agreement with the cluster data and the solar value. \citet{2008ApJ...687.1264M} also noted that there was a slight positive trend with colour index. However, for the sample analysed in this work no strong trend with colour index is found. 

With the exception of the relationship from \citet{1999A&A...348..897L}, there is reasonable agreement between the \citet{1991ApJ...375..722S} and \citet{2008ApJ...687.1264M} relationships and the sample of older cool stars with asteroseismic ages. It should be noted that these previous studies have used various methods to determine ages for their respective samples including isochronal and cluster ages. However, a common issue is the lack of stars older than a gigayear with reliable and accurate ages. This work has improved on this by considering stars with ages determined by asteroseismology, which has proved to be an accurate age-dating method for old field stars. There is some mass scatter at these older ages, as the F-type stars tend to lie below the \citet{2008ApJ...687.1264M} relationship and G-type stars tend to lie above the relationship. This is to be expected as the stellar mass has an effect on the timescale of angular momentum loss. For example, F-type stars are expected to become inactive ($\log R^{'}_{\mathrm{HK}} \sim -5.0$) at approximately 3~Gyr while K-type stars stay active for longer (see Figure 11 from \citet{2008ApJ...687.1264M}).

\begin{figure}
	\centering
    \includegraphics[scale=0.37]{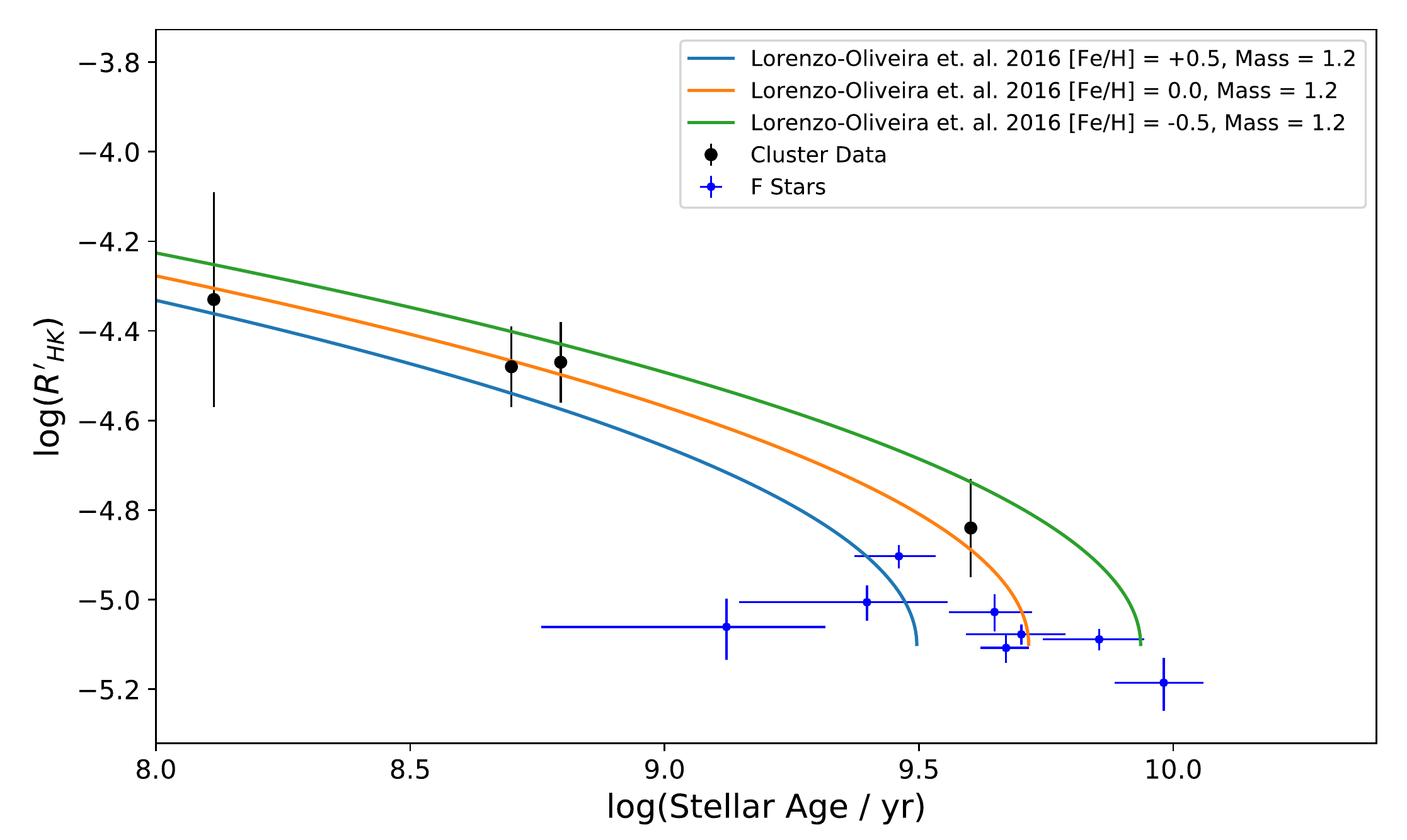}%
    \caption{Plot showing the effect of metallicity on the age-activity relationship. The oldest clusters from \citet{2008ApJ...687.1264M} are shown along with the F stars from the sample of stars analysed in this work.}
	\label{LO_plot}
\end{figure}

Recent work by \citet{2016A&A...594L...3L} has incorporated mass and metallicity into the age-activity relationship and found good agreement with asteroseismic ages, therefore, we also compare our sample of older stars to this relationship. Firstly, we consider variations of the age-mass-metallicity-activity relationship from \citet{2016A&A...594L...3L} and compare these to the stars in our sample. Generally the metallicity has a stronger effect on the relationship given by \citet{2016A&A...594L...3L} than the stellar mass. As an example we show curves for different metallicities and F dwarfs in Figure \ref{LO_plot}; changing the stellar mass by $\pm0.2M_\odot$ has roughly half the effect on the curves than shown here for the metallicities.

\begin{figure}
	\includegraphics[scale=0.36]{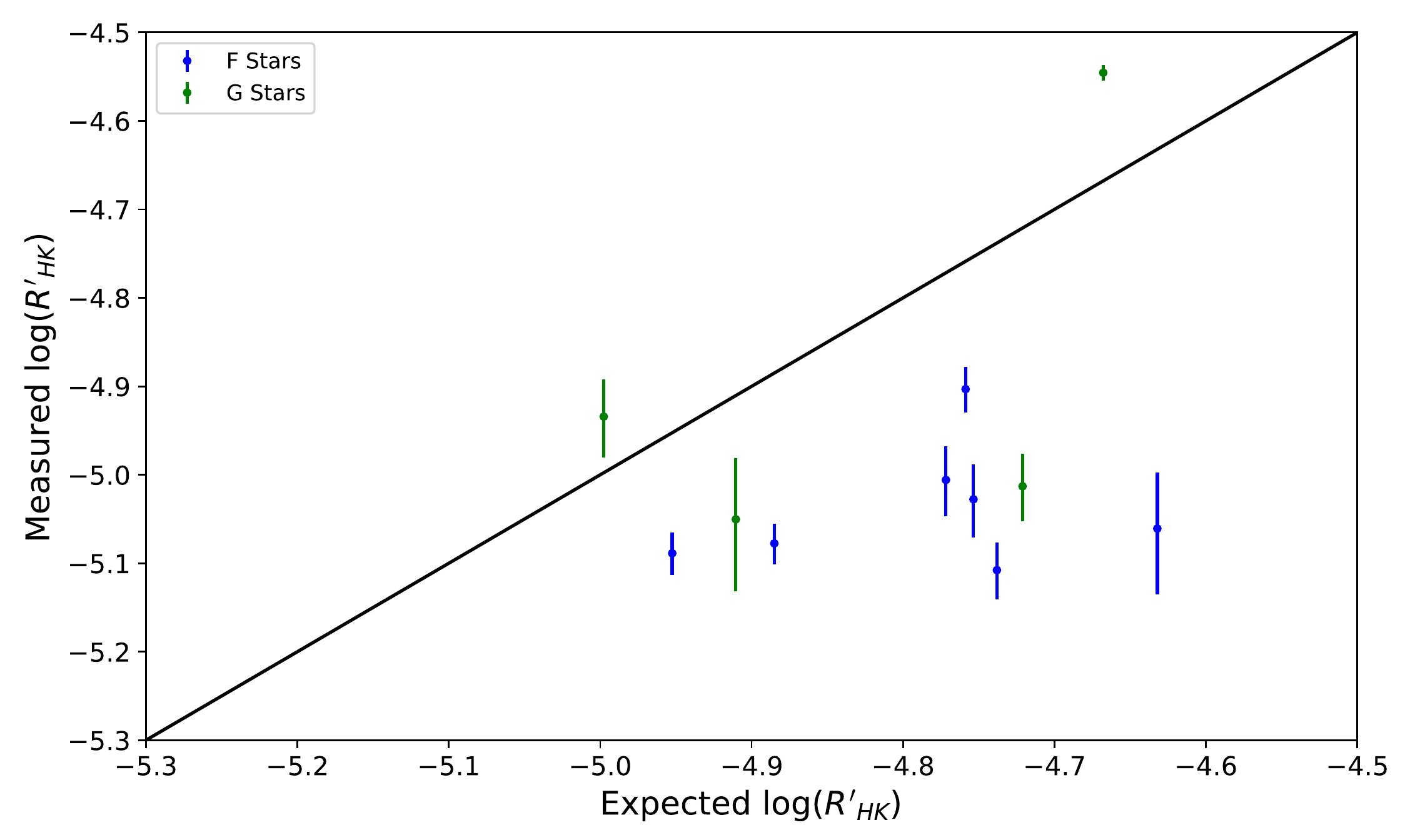}
	\caption{Plot of measured value of the \Rprime indicator from this study as a function of the expected value (as calculated from \citealt{2016A&A...594L...3L}). The black line shows the 1:1 relationship between the two parameters.}
	\centering
	\label{measured_v_expected}
\end{figure}

For our sample of stars, we know the metallicities from the original spectroscopy study \citep{2012MNRAS.423..122B} and the stellar masses from asteroseismology \citep{2014ApJS..210....1C,2017ApJ...835..173S}. We used these parameters along with the asteroseismic age to calculate an expected value for the \Rprime indicator from the age-mass-metallicity-activity relationship \citep{2016A&A...594L...3L}. The measured value of the \Rprime indicator is plotted as a function of the expected value in Figure \ref{measured_v_expected}. Figure \ref{measured_v_expected} shows that for the majority of our sample, the measured value of the \Rprime indicator is less than the expected value from the age-mass-metallicity-activity relationship. This suggests that the majority of our sample stars lie below the predicted age-activity relationship, even after taking mass and metallicity into account. This could potentially be due to a lack of older, inactive stars in the calibration sample of \citet{2016A&A...594L...3L} or a real discrepancy between older stars and predicted relationships.

\subsection{Spin-down in the context of the observed data}
\label{spindown}

In recent years, the availability of stellar ages from asteroseismology has fuelled new investigations into the nature of rotation and magnetic activity--age relationships for old stars \citep{2016Natur.529..181V,2017MNRAS.471.1012B}. \citet{2016Natur.529..181V} investigated the rotation--age relationship using data from \textit{Kepler} and asteroseismic ages. They found anomalously high rotation rates for older stars and suggested that this was due to weakened magnetic braking. The investigation of magnetic activity levels for similarly old stars shows a different trend: \citet{2017MNRAS.471.1012B} investigated the age-activity relationship by combining X-ray observations with ages from a variety of age-determination methods (including asteroseismology). They found that the X-ray luminosity decays more rapidly for older stars compared to the younger cluster data. One possible explanation for this steepening of the age-activity relationship is due to a change in the activity-rotation relationship; this would mean that after a certain stellar age a decline in magnetic activity does no longer go hand in hand with rotational spin-down, as seen for younger stars.

Interestingly, our sample investigated in this work has a very similar spectral range as the stars investigated for rotational behaviour by \citet{2016Natur.529..181V}, namely mostly late F stars and some G stars. However, our sample does not show abnormally high values of the \Rprime indicator that may be associated with weakened magnetic braking at these older ages \citep{2016Natur.529..181V}. Both for the G-type stars and the cooler F-type stars which still have a convective envelope, our data indeed shows that magnetic activity is still decreasing even at old stellar ages.

Possible explanations for these observations may lie in the detectability of various properties used in age-rotation-activity, as well as in genuine physical properties of stars. On the detectability side, the question is which stars allow us to measure their age asteroseismically, to measure their magnetic activity level from chromospheric (or coronal) emission, and to detect rotational modulation in their light curves.

To measure an asteroseismic age, given a suitable light curve, it has been shown that high magnetic activity makes the detection of asteroseismic modes more difficult; this seems to be not purely a signal analysis issue, but possibly a genuine suppression of asteroseismic oscillations by magnetic activity \citep{2011ApJ...732L...5C}. This means that when we study stars with measured asteroseismic ages, we are automatically selecting a sample biased towards low-activity stars. We think that this does not have a significant effect on the type of old, low-activity stars we are studying here, because the suppression of asteroseismic modes was reported for stars with higher variability and activity level.

To measure chromospheric or coronal emission as an indicator for magnetic activity, a flux measurement is performed. For chromospheric lines this is typically done by a relative measurement of the flux in an active line compared to some surrounding continuum, and for coronal emission the absolute flux can be measured (but is often also put into context with the absolute bolometric luminosity of the star). Upper limits for coronal X-ray emission are derived from photon statistics. For chromospheric activity the absence of an activity-related flux excess in relevant lines can be determined, since the purely photospheric component in e.g.\ the calcium II H\&K lines is well studied \citep{1984ApJ...279..763N}. Therefore, detecting that a star has a very low level of stellar activity is observationally relatively straightforward.

To measure stellar rotation from light curve modulation, the star needs to cooperate by displaying a sufficient difference in starspot (or plage) coverage on its hemispheres to produce a detectable brightness modulation. Stars with particularly low activity display lower variability in broadband optical light curves, making a period determination difficult. Furthermore, deriving an upper limit on rotation period from a light curve is not straightforward, as it involves both the sampling of a period and the amplitude of the modulation. We would therefore argue that the detectabilty of periods -- somewhat counterintuitively, as they are the physically more fundamental measurement related to spin-down -- is actually more vulnerable to selection effects than the measurement of stellar activity levels.

Addressing the question of the age--rotation--activity discrepancies from the physical side, it is also very well possible that the stalling spin-down and strongly decreasing activity are both real. For stars with old main-sequence ages we have very little information available about their spin-down, as appropriately old stellar clusters are not easily accessible for rotation and activity measurements, or even asteroseismic age determinations for individual cluster stars. 

Therefore, a possible physical explanations is that the evolution of rotation with age and activity with age do not go hand in hand any longer after the star reaches an old age, as has been suggested by \citet{2017MNRAS.471.1012B}. This would mean a change in the rate of rotational braking. Such changes in braking have also been suggested for young stars in order to explain how they move from the fast rotation branch to the slow rotation branch \citep{2015ApJ...813...40G, 2018ApJ...862...90G}; specifically, \citet{2016A&A...595A.110G} suggest that a switch of the stellar magnetic field from a high-order multipole to a low-order one dramatically increases the magnetic braking mediated by the stellar wind. It is possible that for old stars yet another change in magnetic braking takes place, which switches the star to marginal braking again. Another study investigated the stellar wind in simulations, albeit using models with simplifying assumptions; this suggests decreasing mass-loss rates for Sun like stars around ages of ca.\ 2\,Gyr \citep{2018MNRAS.476.2465A}. 

A potential physical pathway for the observations of stalling  rotational spin-down and decreasing magnetic activity may therefore be that a switch in magnetic morphology decreases the stellar wind along with the chromospheric and coronal brightness of active regions, causing slower removal of rotational momentum, which leads to longer rotation periods as seen in the \textit{Kepler} observations.

\subsection{Chromospheric activity as an age indicator for old stars}

In recent years, there has been some debate over the use of chromospheric emission as an age indicator, particularly for stars older than a gigayear. \citet{2013A&A...551L...8P} presented an L-shaped chromospheric activity versus age diagram and suggested that the use of chromospheric emission as an age indicator is limited to stars younger than $\approx 1.5$ Gyr. However, recent work from \citet{2016A&A...594L...3L} incorporated mass and metallicity into the age-activity relationship as discussed in Section \ref{previous_relationships}. In their study, they were also able to reproduce the lack of chromospheric evolution (Figure 2 within) by following the same sample selection procedure as \citet{2013A&A...551L...8P}. \citet{2016A&A...594L...3L} found that there was a constant metallicity dispersion along the age axis, which can be understood as an additional source of scatter in the age-activity relationship. They also found that as they considered younger stars, the sample started to become more biased towards higher mass stars.

Our sample of stars shows that there is still correlation between chromospheric activity and age for a given spectral type at ages older than a gigayear. Therefore, the possibility of constructing an age--activity relationship is still viable at these older ages provided that the mass and metallicity are taken into account. This is more important for older stars as the age at which they become inactive ($\log R^{'}_{\mathrm{HK}} \sim -5.0$) is dependent on the stellar mass and as demonstrated by \citet{2016A&A...594L...3L}, the metallicity also has an effect on the shape of the age--activity relationship. For example, \citet{2018A&A...619A..73L} considered a sample of solar twins and found that the age--activity relationship remained statistically significant up to $\sim 6-7$~Gyr. Therefore, tight sample constraints can allow for age--activity relationships even at older ages. However, a generic age-activity relationship for all spectral types and metallicities using the \Rprime indicator does not seem possible.

\section{Conclusions}
\label{conclusions}
In this work we have presented a sample of 26 cool stars with asteroseismic ages along with calculated values for the \Rprime indicator in an attempt to calibrate the age-activity relationship beyond a gigayear. Relationships were found between the S index calculated in the \narval and \esp spectrographs and \smw using stars from \citet{1991ApJS...76..383D}. This allowed for a conversion to \smw and the original method by \citet{1984ApJ...279..763N} was used to calculate the \Rprime indicator. As expected, the early F-type stars that lie above the Kraft break do not display a correlation between age and activity. However, they do display chromospheric activity with a range of \Rprime values primarily between $-5.1< \mathrm{log}(R^{'}_{HK})< -4.8$. We also find that for the cool star part of the sample (i.e. $T_{eff} < 6200$ K), for a given spectral type, there is correlation between the chromospheric activity and age.

We compared our sample of old stars to previous age-activity relationships and find reasonable agreement with the \citet{2008ApJ...687.1264M} relationship with some scatter due to stellar mass. This is consistent with F-type stars becoming inactive at younger ages than G-type stars. We also compared our sample to the age-mass-metallicity-activity relationship from \citet{2016A&A...594L...3L} to see the effects of mass and/or metallicity to the age--activity relationship at older ages. We found that even after taking stellar mass and metallicity into account, most of our sample stars display lower activity than predicted. This would suggest either a lack of older, inactive stars in the calibration of the \citet{2016A&A...594L...3L} relationship or a real discrepancy between older stars and predicted relationships.

The very low activity of the studied sample is in line with previous work on coronal X-ray activity for a sample of old cool stars with a variety of age-determination methods \citep{2017MNRAS.471.1012B}. We discussed the observed discrepancy with moderately fast rotation periods for similarly old stars with asteroseismic ages \citep{2016Natur.529..181V} in the context of both detectability questions for the physical quantities involves, and a genuine departure from the parallel decrease in rotation rate and magnetic activity which may be present in old main-sequence stars.

Our sample of stars also allows for discussion on the suitability of chromospheric emission as an age indicator for stars older than a gigayear. Given the correlation seen for a given spectral type in our sample, it would suggest that with careful constraints on mass and metallicity it may be possible to calibrate an age-activity relationship for stars with a narrow range of stellar parameters.

\section*{Acknowledgements}
We thank the anonymous reviewer for their comments and suggestions, which added significantly to the clarity and discussion of the results. This research made use of public databases hosted by SIMBAD, maintained by CDS, Strasbourg, France. RB acknowledges funding from DfE. CAW gratefully acknowledges support from STFC grant ST/P000312/1. Funding for the Stellar Astrophysics Centre is provided by The Danish National Research Foundation (Grant agreement No.~DNRF106). V.S.A. acknowledges support from the Independent Research Fund Denmark (Research grant 7027-00096B).

%%%%%%%%%%%%%%%%%%%% REFERENCES %%%%%%%%%%%%%%%%%%

\bibliographystyle{mnras}
\bibliography{references}

\begin{thebibliography}{}
\makeatletter
\relax
\def\mn@urlcharsother{\let\do\@makeother \do\$\do\&\do\#\do\^\do\_\do\%\do\~}
\def\mn@doi{\begingroup\mn@urlcharsother \@ifnextchar [ {\mn@doi@}
  {\mn@doi@[]}}
\def\mn@doi@[#1]#2{\def\@tempa{#1}\ifx\@tempa\@empty \href
  {http://dx.doi.org/#2} {doi:#2}\else \href {http://dx.doi.org/#2} {#1}\fi
  \endgroup}
\def\mn@eprint#1#2{\mn@eprint@#1:#2::\@nil}
\def\mn@eprint@arXiv#1{\href {http://arxiv.org/abs/#1} {{\tt arXiv:#1}}}
\def\mn@eprint@dblp#1{\href {http://dblp.uni-trier.de/rec/bibtex/#1.xml}
  {dblp:#1}}
\def\mn@eprint@#1:#2:#3:#4\@nil{\def\@tempa {#1}\def\@tempb {#2}\def\@tempc
  {#3}\ifx \@tempc \@empty \let \@tempc \@tempb \let \@tempb \@tempa \fi \ifx
  \@tempb \@empty \def\@tempb {arXiv}\fi \@ifundefined
  {mn@eprint@\@tempb}{\@tempb:\@tempc}{\expandafter \expandafter \csname
  mn@eprint@\@tempb\endcsname \expandafter{\@tempc}}}

\bibitem[\protect\citeauthoryear{{Auri{\`e}re}}{{Auri{\`e}re}}{2003}]{2003EAS.....9..105A}
{Auri{\`e}re} M.,  2003, in {Arnaud} J.,  {Meunier} N.,  eds,  EAS Publications
  Series Vol. 9, EAS Publications Series. p.~105

\bibitem[\protect\citeauthoryear{{Bahcall}, {Pinsonneault}  \&
  {Wasserburg}}{{Bahcall} et~al.}{1995}]{1995RvMP...67..781B}
{Bahcall} J.~N.,  {Pinsonneault} M.~H.,   {Wasserburg} G.~J.,  1995, \mn@doi
  [Reviews of Modern Physics] {10.1103/RevModPhys.67.781}, \href
  {http://adsabs.harvard.edu/abs/1995RvMP...67..781B} {67, 781}

\bibitem[\protect\citeauthoryear{{Baliunas} et~al.,}{{Baliunas}
  et~al.}{1995}]{1995ApJ...438..269B}
{Baliunas} S.~L.,  et~al., 1995, \mn@doi [\apj] {10.1086/175072}, \href
  {http://adsabs.harvard.edu/abs/1995ApJ...438..269B} {438, 269}

\bibitem[\protect\citeauthoryear{{Barnes}, {Weingrill}, {Fritzewski},
  {Strassmeier}  \& {Platais}}{{Barnes} et~al.}{2016}]{2016ApJ...823...16B}
{Barnes} S.~A.,  {Weingrill} J.,  {Fritzewski} D.,  {Strassmeier} K.~G.,
  {Platais} I.,  2016, \mn@doi [\apj] {10.3847/0004-637X/823/1/16}, \href
  {http://adsabs.harvard.edu/abs/2016ApJ...823...16B} {823, 16}

\bibitem[\protect\citeauthoryear{{Bellini} et~al.,}{{Bellini}
  et~al.}{2010}]{2010A&A...513A..50B}
{Bellini} A.,  et~al., 2010, \mn@doi [\aap] {10.1051/0004-6361/200913721},
  \href {http://adsabs.harvard.edu/abs/2010A%26A...513A..50B} {513, A50}

\bibitem[\protect\citeauthoryear{{Booth}, {Poppenhaeger}, {Watson}, {Silva
  Aguirre}  \& {Wolk}}{{Booth} et~al.}{2017}]{2017MNRAS.471.1012B}
{Booth} R.~S.,  {Poppenhaeger} K.,  {Watson} C.~A.,  {Silva Aguirre} V.,
  {Wolk} S.~J.,  2017, \mn@doi [\mnras] {10.1093/mnras/stx1630}, \href
  {http://adsabs.harvard.edu/abs/2017MNRAS.471.1012B} {471, 1012}

\bibitem[\protect\citeauthoryear{{Bruntt} et~al.,}{{Bruntt}
  et~al.}{2012}]{2012MNRAS.423..122B}
{Bruntt} H.,  et~al., 2012, \mn@doi [\mnras]
  {10.1111/j.1365-2966.2012.20686.x}, \href
  {http://adsabs.harvard.edu/abs/2012MNRAS.423..122B} {423, 122}

\bibitem[\protect\citeauthoryear{{Chaplin} \& {Miglio}}{{Chaplin} \&
  {Miglio}}{2013}]{2013ARA&A..51..353C}
{Chaplin} W.~J.,  {Miglio} A.,  2013, \mn@doi [\araa]
  {10.1146/annurev-astro-082812-140938}, \href
  {http://adsabs.harvard.edu/abs/2013ARA%26A..51..353C} {51, 353}

\bibitem[\protect\citeauthoryear{{Chaplin} et~al.,}{{Chaplin}
  et~al.}{2011}]{2011ApJ...732L...5C}
{Chaplin} W.~J.,  et~al., 2011, \mn@doi [\apjl] {10.1088/2041-8205/732/1/L5},
  \href {http://adsabs.harvard.edu/abs/2011ApJ...732L...5C} {732, L5}

\bibitem[\protect\citeauthoryear{{Chaplin} et~al.,}{{Chaplin}
  et~al.}{2014}]{2014ApJS..210....1C}
{Chaplin} W.~J.,  et~al., 2014, \mn@doi [\apjs] {10.1088/0067-0049/210/1/1},
  \href {http://adsabs.harvard.edu/abs/2014ApJS..210....1C} {210, 1}

\bibitem[\protect\citeauthoryear{{Cuntz}, {Saar}  \& {Musielak}}{{Cuntz}
  et~al.}{2000}]{2000ApJ...533L.151C}
{Cuntz} M.,  {Saar} S.~H.,   {Musielak} Z.~E.,  2000, \mn@doi [\apjl]
  {10.1086/312609}, \href {http://adsabs.harvard.edu/abs/2000ApJ...533L.151C}
  {533, L151}

\bibitem[\protect\citeauthoryear{{Curtis}}{{Curtis}}{2017}]{2017AJ....153..275C}
{Curtis} J.~L.,  2017, \mn@doi [\aj] {10.3847/1538-3881/aa72e9}, \href
  {https://ui.adsabs.harvard.edu/abs/2017AJ....153..275C} {153, 275}

\bibitem[\protect\citeauthoryear{{Demarque}, {Green}  \& {Guenther}}{{Demarque}
  et~al.}{1992}]{1992AJ....103..151D}
{Demarque} P.,  {Green} E.~M.,   {Guenther} D.~B.,  1992, \mn@doi [\aj]
  {10.1086/116049}, \href {http://adsabs.harvard.edu/abs/1992AJ....103..151D}
  {103, 151}

\bibitem[\protect\citeauthoryear{{Donati}, {Catala}, {Landstreet}  \&
  {Petit}}{{Donati} et~al.}{2006}]{2006ASPC..358..362D}
{Donati} J.-F.,  {Catala} C.,  {Landstreet} J.~D.,   {Petit} P.,  2006, in
  {Casini} R.,  {Lites} B.~W.,  eds,  Astronomical Society of the Pacific
  Conference Series Vol. 358, Astronomical Society of the Pacific Conference
  Series. p.~362

\bibitem[\protect\citeauthoryear{{Duncan} et~al.,}{{Duncan}
  et~al.}{1991}]{1991ApJS...76..383D}
{Duncan} D.~K.,  et~al., 1991, \mn@doi [\apjs] {10.1086/191572}, \href
  {http://adsabs.harvard.edu/abs/1991ApJS...76..383D} {76, 383}

\bibitem[\protect\citeauthoryear{{Egeland}, {Soon}, {Baliunas}, {Hall},
  {Pevtsov}  \& {Bertello}}{{Egeland} et~al.}{2017}]{2017ApJ...835...25E}
{Egeland} R.,  {Soon} W.,  {Baliunas} S.,  {Hall} J.~C.,  {Pevtsov} A.~A.,
  {Bertello} L.,  2017, \mn@doi [\apj] {10.3847/1538-4357/835/1/25}, \href
  {http://adsabs.harvard.edu/abs/2017ApJ...835...25E} {835, 25}

\bibitem[\protect\citeauthoryear{{Garraffo}, {Drake}  \& {Cohen}}{{Garraffo}
  et~al.}{2015}]{2015ApJ...813...40G}
{Garraffo} C.,  {Drake} J.~J.,   {Cohen} O.,  2015, \mn@doi [\apj]
  {10.1088/0004-637X/813/1/40}, \href
  {http://adsabs.harvard.edu/abs/2015ApJ...813...40G} {813, 40}

\bibitem[\protect\citeauthoryear{{Garraffo}, {Drake}  \& {Cohen}}{{Garraffo}
  et~al.}{2016}]{2016A&A...595A.110G}
{Garraffo} C.,  {Drake} J.~J.,   {Cohen} O.,  2016, \mn@doi [\aap]
  {10.1051/0004-6361/201628367}, \href
  {http://adsabs.harvard.edu/abs/2016A%26A...595A.110G} {595, A110}

\bibitem[\protect\citeauthoryear{{Garraffo} et~al.,}{{Garraffo}
  et~al.}{2018}]{2018ApJ...862...90G}
{Garraffo} C.,  et~al., 2018, \mn@doi [\apj] {10.3847/1538-4357/aace5d}, \href
  {http://adsabs.harvard.edu/abs/2018ApJ...862...90G} {862, 90}

\bibitem[\protect\citeauthoryear{{Giampapa}, {Hall}, {Radick}  \&
  {Baliunas}}{{Giampapa} et~al.}{2006}]{2006ApJ...651..444G}
{Giampapa} M.~S.,  {Hall} J.~C.,  {Radick} R.~R.,   {Baliunas} S.~L.,  2006,
  \mn@doi [\apj] {10.1086/507624}, \href
  {http://adsabs.harvard.edu/abs/2006ApJ...651..444G} {651, 444}

\bibitem[\protect\citeauthoryear{Gray}{Gray}{2005}]{CBO9781316036570A207}
Gray D.~F.,  2005, in , The Observation and Analysis of Stellar Photospheres,
  third edn, Cambridge University Press, pp 365--383

\bibitem[\protect\citeauthoryear{{Hall}, {Lockwood}  \& {Skiff}}{{Hall}
  et~al.}{2007}]{2007AJ....133..862H}
{Hall} J.~C.,  {Lockwood} G.~W.,   {Skiff} B.~A.,  2007, \mn@doi [\aj]
  {10.1086/510356}, \href {http://adsabs.harvard.edu/abs/2007AJ....133..862H}
  {133, 862}

\bibitem[\protect\citeauthoryear{{Hall}, {Henry}, {Lockwood}, {Skiff}  \&
  {Saar}}{{Hall} et~al.}{2009}]{2009AJ....138..312H}
{Hall} J.~C.,  {Henry} G.~W.,  {Lockwood} G.~W.,  {Skiff} B.~A.,   {Saar}
  S.~H.,  2009, \mn@doi [\aj] {10.1088/0004-6256/138/1/312}, \href
  {http://adsabs.harvard.edu/abs/2009AJ....138..312H} {138, 312}

\bibitem[\protect\citeauthoryear{{Kraft}}{{Kraft}}{1967}]{1967ApJ...150..551K}
{Kraft} R.~P.,  1967, \mn@doi [\apj] {10.1086/149359}, \href
  {https://ui.adsabs.harvard.edu/abs/1967ApJ...150..551K} {150, 551}

\bibitem[\protect\citeauthoryear{{Lachaume}, {Dominik}, {Lanz}  \&
  {Habing}}{{Lachaume} et~al.}{1999}]{1999A&A...348..897L}
{Lachaume} R.,  {Dominik} C.,  {Lanz} T.,   {Habing} H.~J.,  1999, \aap, \href
  {http://adsabs.harvard.edu/abs/1999A%26A...348..897L} {348, 897}

\bibitem[\protect\citeauthoryear{{Lallement}, {Babusiaux}, {Vergely}, {Katz},
  {Arenou}, {Valette}, {Hottier}  \& {Capitanio}}{{Lallement}
  et~al.}{2019}]{2019A&A...625A.135L}
{Lallement} R.,  {Babusiaux} C.,  {Vergely} J.~L.,  {Katz} D.,  {Arenou} F.,
  {Valette} B.,  {Hottier} C.,   {Capitanio} L.,  2019, \mn@doi [\aap]
  {10.1051/0004-6361/201834695}, \href
  {https://ui.adsabs.harvard.edu/abs/2019A&A...625A.135L} {625, A135}

\bibitem[\protect\citeauthoryear{{Lorenzo-Oliveira}, {Porto de Mello}  \&
  {Schiavon}}{{Lorenzo-Oliveira} et~al.}{2016}]{2016A&A...594L...3L}
{Lorenzo-Oliveira} D.,  {Porto de Mello} G.~F.,   {Schiavon} R.~P.,  2016,
  \mn@doi [\aap] {10.1051/0004-6361/201629233}, \href
  {http://adsabs.harvard.edu/abs/2016A%26A...594L...3L} {594, L3}

\bibitem[\protect\citeauthoryear{{Lorenzo-Oliveira} et~al.,}{{Lorenzo-Oliveira}
  et~al.}{2018}]{2018A&A...619A..73L}
{Lorenzo-Oliveira} D.,  et~al., 2018, \mn@doi [\aap]
  {10.1051/0004-6361/201629294}, \href
  {https://ui.adsabs.harvard.edu/abs/2018A&A...619A..73L} {619, A73}

\bibitem[\protect\citeauthoryear{{Lovis} et~al.,}{{Lovis}
  et~al.}{2011}]{2011arXiv1107.5325L}
{Lovis} C.,  et~al., 2011, preprint, \href
  {http://adsabs.harvard.edu/abs/2011arXiv1107.5325L} {} (\mn@eprint {arXiv}
  {1107.5325})

\bibitem[\protect\citeauthoryear{{Mamajek} \& {Hillenbrand}}{{Mamajek} \&
  {Hillenbrand}}{2008}]{2008ApJ...687.1264M}
{Mamajek} E.~E.,  {Hillenbrand} L.~A.,  2008, \mn@doi [\apj] {10.1086/591785},
  \href {http://adsabs.harvard.edu/abs/2008ApJ...687.1264M} {687, 1264}

\bibitem[\protect\citeauthoryear{{Marcy} et~al.,}{{Marcy}
  et~al.}{2014}]{2014ApJS..210...20M}
{Marcy} G.~W.,  et~al., 2014, \mn@doi [\apjs] {10.1088/0067-0049/210/2/20},
  \href {http://adsabs.harvard.edu/abs/2014ApJS..210...20M} {210, 20}

\bibitem[\protect\citeauthoryear{{Mathur} et~al.,}{{Mathur}
  et~al.}{2012}]{2012ApJ...749..152M}
{Mathur} S.,  et~al., 2012, \mn@doi [\apj] {10.1088/0004-637X/749/2/152}, \href
  {http://adsabs.harvard.edu/abs/2012ApJ...749..152M} {749, 152}

\bibitem[\protect\citeauthoryear{{Metcalfe} et~al.,}{{Metcalfe}
  et~al.}{2012}]{2012ApJ...748L..10M}
{Metcalfe} T.~S.,  et~al., 2012, \mn@doi [\apjl] {10.1088/2041-8205/748/1/L10},
  \href {http://adsabs.harvard.edu/abs/2012ApJ...748L..10M} {748, L10}

\bibitem[\protect\citeauthoryear{{Noyes}, {Hartmann}, {Baliunas}, {Duncan}  \&
  {Vaughan}}{{Noyes} et~al.}{1984}]{1984ApJ...279..763N}
{Noyes} R.~W.,  {Hartmann} L.~W.,  {Baliunas} S.~L.,  {Duncan} D.~K.,
  {Vaughan} A.~H.,  1984, \mn@doi [\apj] {10.1086/161945}, \href
  {http://adsabs.harvard.edu/abs/1984ApJ...279..763N} {279, 763}

\bibitem[\protect\citeauthoryear{{{\'O} Fionnag{\'a}in} \& Vidotto}{{{\'O}
  Fionnag{\'a}in} \& Vidotto}{2018}]{2018MNRAS.476.2465A}
{{\'O} Fionnag{\'a}in} D.,  Vidotto A.~A.,  2018, \mn@doi [\mnras]
  {10.1093/mnras/sty394}, 476, 2465

\bibitem[\protect\citeauthoryear{{Pace}}{{Pace}}{2013}]{2013A&A...551L...8P}
{Pace} G.,  2013, \mn@doi [\aap] {10.1051/0004-6361/201220364}, \href
  {http://adsabs.harvard.edu/abs/2013A%26A...551L...8P} {551, L8}

\bibitem[\protect\citeauthoryear{{Parker}}{{Parker}}{1955}]{1955ApJ...122..293P}
{Parker} E.~N.,  1955, \mn@doi [\apj] {10.1086/146087}, \href
  {http://adsabs.harvard.edu/abs/1955ApJ...122..293P} {122, 293}

\bibitem[\protect\citeauthoryear{{Pillitteri}, {Maggio}, {Micela}, {Sciortino},
  {Wolk}  \& {Matsakos}}{{Pillitteri} et~al.}{2015}]{2015ApJ...805...52P}
{Pillitteri} I.,  {Maggio} A.,  {Micela} G.,  {Sciortino} S.,  {Wolk} S.~J.,
  {Matsakos} T.,  2015, \mn@doi [\apj] {10.1088/0004-637X/805/1/52}, \href
  {http://adsabs.harvard.edu/abs/2015ApJ...805...52P} {805, 52}

\bibitem[\protect\citeauthoryear{{Poppenhaeger} \& {Wolk}}{{Poppenhaeger} \&
  {Wolk}}{2014}]{2014A&A...565L...1P}
{Poppenhaeger} K.,  {Wolk} S.~J.,  2014, \mn@doi [\aap]
  {10.1051/0004-6361/201423454}, \href
  {http://adsabs.harvard.edu/abs/2014A%26A...565L...1P} {565, L1}

\bibitem[\protect\citeauthoryear{{Robrade} \& {Schmitt}}{{Robrade} \&
  {Schmitt}}{2009}]{2009A&A...497..511R}
{Robrade} J.,  {Schmitt} J.~H.~M.~M.,  2009, \mn@doi [\aap]
  {10.1051/0004-6361/200811348}, \href
  {http://adsabs.harvard.edu/abs/2009A%26A...497..511R} {497, 511}

\bibitem[\protect\citeauthoryear{{Rocha-Pinto} \& {Maciel}}{{Rocha-Pinto} \&
  {Maciel}}{1998}]{1998MNRAS.298..332R}
{Rocha-Pinto} H.~J.,  {Maciel} W.~J.,  1998, \mn@doi [\mnras]
  {10.1046/j.1365-8711.1998.01597.x}, \href
  {http://adsabs.harvard.edu/abs/1998MNRAS.298..332R} {298, 332}

\bibitem[\protect\citeauthoryear{{Schatzman}}{{Schatzman}}{1962}]{1962AnAp...25...18S}
{Schatzman} E.,  1962, Annales d'Astrophysique, \href
  {http://adsabs.harvard.edu/abs/1962AnAp...25...18S} {25, 18}

\bibitem[\protect\citeauthoryear{{Silva Aguirre} et~al.,}{{Silva Aguirre}
  et~al.}{2015}]{2015MNRAS.452.2127S}
{Silva Aguirre} V.,  et~al., 2015, \mn@doi [\mnras] {10.1093/mnras/stv1388},
  \href {http://adsabs.harvard.edu/abs/2015MNRAS.452.2127S} {452, 2127}

\bibitem[\protect\citeauthoryear{{Silva Aguirre} et~al.,}{{Silva Aguirre}
  et~al.}{2017}]{2017ApJ...835..173S}
{Silva Aguirre} V.,  et~al., 2017, \mn@doi [\apj]
  {10.3847/1538-4357/835/2/173}, \href
  {http://adsabs.harvard.edu/abs/2017ApJ...835..173S} {835, 173}

\bibitem[\protect\citeauthoryear{{Skumanich}}{{Skumanich}}{1972}]{1972ApJ...171..565S}
{Skumanich} A.,  1972, \mn@doi [\apj] {10.1086/151310}, \href
  {http://adsabs.harvard.edu/abs/1972ApJ...171..565S} {171, 565}

\bibitem[\protect\citeauthoryear{{Soderblom}, {Duncan}  \&
  {Johnson}}{{Soderblom} et~al.}{1991}]{1991ApJ...375..722S}
{Soderblom} D.~R.,  {Duncan} D.~K.,   {Johnson} D.~R.~H.,  1991, \mn@doi [\apj]
  {10.1086/170238}, \href {http://adsabs.harvard.edu/abs/1991ApJ...375..722S}
  {375, 722}

\bibitem[\protect\citeauthoryear{{VandenBerg} \& {Stetson}}{{VandenBerg} \&
  {Stetson}}{2004}]{2004PASP..116..997V}
{VandenBerg} D.~A.,  {Stetson} P.~B.,  2004, \mn@doi [\pasp] {10.1086/426340},
  \href {http://adsabs.harvard.edu/abs/2004PASP..116..997V} {116, 997}

\bibitem[\protect\citeauthoryear{{Wenger} et~al.,}{{Wenger}
  et~al.}{2000}]{2000A&AS..143....9W}
{Wenger} M.,  et~al., 2000, \mn@doi [A{\&}AS] {10.1051/aas:2000332}, \href
  {http://adsabs.harvard.edu/abs/2000A%26AS..143....9W} {143, 9}

\bibitem[\protect\citeauthoryear{{Wilson}}{{Wilson}}{1968}]{1968ApJ...153..221W}
{Wilson} O.~C.,  1968, \mn@doi [\apj] {10.1086/149652}, \href
  {http://adsabs.harvard.edu/abs/1968ApJ...153..221W} {153, 221}

\bibitem[\protect\citeauthoryear{{Wilson}}{{Wilson}}{1978}]{1978ApJ...226..379W}
{Wilson} O.~C.,  1978, \mn@doi [\apj] {10.1086/156618}, \href
  {http://adsabs.harvard.edu/abs/1978ApJ...226..379W} {226, 379}

\bibitem[\protect\citeauthoryear{{Wolff} \& {Simon}}{{Wolff} \&
  {Simon}}{1997}]{1997PASP..109..759W}
{Wolff} S.,  {Simon} T.,  1997, \mn@doi [\pasp] {10.1086/133942}, \href
  {https://ui.adsabs.harvard.edu/abs/1997PASP..109..759W} {109, 759}

\bibitem[\protect\citeauthoryear{{Zorec} \& {Royer}}{{Zorec} \&
  {Royer}}{2012}]{2012A&A...537A.120Z}
{Zorec} J.,  {Royer} F.,  2012, \mn@doi [\aap] {10.1051/0004-6361/201117691},
  \href {https://ui.adsabs.harvard.edu/abs/2012A&A...537A.120Z} {537, A120}

\bibitem[\protect\citeauthoryear{{van Saders}, {Ceillier}, {Metcalfe}, {Silva
  Aguirre}, {Pinsonneault}, {Garc{\'{\i}}a}, {Mathur}  \& {Davies}}{{van
  Saders} et~al.}{2016}]{2016Natur.529..181V}
{van Saders} J.~L.,  {Ceillier} T.,  {Metcalfe} T.~S.,  {Silva Aguirre} V.,
  {Pinsonneault} M.~H.,  {Garc{\'{\i}}a} R.~A.,  {Mathur} S.,   {Davies} G.~R.,
   2016, \mn@doi [\nat] {10.1038/nature16168}, \href
  {http://adsabs.harvard.edu/abs/2016Natur.529..181V} {529, 181}

\makeatother
\end{thebibliography}

\clearpage
\newpage

\onecolumn
\appendix

\clearpage
\newpage
\begin{landscape}
\section{\texorpdfstring{\Rprime}{Rprime} and age results}
\label{calcium_results}

%Cool stars table
\begin{minipage}{\linewidth}
\centering
%\begin{small}
\bgroup
\def\arraystretch{1.4}
{
\begin{tabular}{llcccccccccl}
\hline
Star Name    & [Fe/H] & M / $M_{\odot}$ & $B-V$  & $T_{eff}$ / K & $log(g)$ & Age / Gyr & 	S Index      & \smw  & $log(R^{'}_{HK})$ & Spectrograph \\
\hline
KIC 10462940 & 0.10 & $1.20^{+0.07}_{-0.06}$ & 0.56 & 6154 & $4.32$   & $2.50^{+1.10}_{-1.10}$ \footnote{Age and mass taken from \citet{2014ApJS..210....1C}\label{C14}}      & 0.0071 & 0.1540 & $-5.01^{+0.04}_{-0.04}$ & \narval       \\
KIC 9139151  & 0.11 & $1.18^{+0.04}_{-0.05}$ & 0.51 & 6125 & $4.38$   & $1.32^{+0.75}_{-0.75}$ \footnote{Age and mass taken from \citet{2017ApJ...835..173S}\label{A17}}      & 0.0068 & 0.1473 & $-5.06^{+0.06}_{-0.07}$ & \esp          \\
KIC 10454113 & -0.06 & $1.17^{+0.02}_{-0.03}$ & 0.52 & 6120 & $4.31$   & $2.89^{+0.53}_{-0.53}$ \footref{A17}      & 0.0082 & 0.1698 & $-4.89^{+0.03}_{-0.03}$ & \esp          \\
KIC 10454113 & -0.06 & $1.17^{+0.02}_{-0.03}$ & 0.52 & 6120 & $4.31$   & $2.89^{+0.53}_{-0.53}$ \footref{A17}      & 0.0078 & 0.1678 & $-4.90^{+0.03}_{-0.03}$ & \narval       \\
KIC 8694723  & -0.59 & $1.14^{+0.02}_{-0.02}$ & 0.46 & 6120 & $4.10$   & $4.69^{+0.51}_{-0.51}$ \footref{A17}      & 0.0063 & 0.1398 & $-5.11^{+0.03}_{-0.03}$ & \narval       \\
KIC 8394589  & -0.36 & $1.04^{+0.04}_{-0.03}$ & 0.55 & 6114 & $4.32$   & $4.45^{+0.83}_{-0.83}$ \footref{A17}      & 0.0069 & 0.1507 & $-5.03^{+0.04}_{-0.04}$ & \narval       \\
KIC 10963065 & -0.2 & $0.99^{+0.06}_{-0.06}$ & 0.51 & 6060 & $4.29$   & $7.15^{+1.61}_{-1.61}$ \footref{A17}      & 0.0064 & 0.1423 & $-5.09^{+0.02}_{-0.02}$ & \narval       \\
KIC 6106415  & -0.09 & $1.07^{+0.05}_{-0.04}$ & 0.55 & 5990 & $4.31$   & $5.03^{+1.12}_{-1.12}$ \footref{A17}      & 0.0066 & 0.1447 & $-5.08^{+0.02}_{-0.02}$ & \narval       \\
KIC 6116048  & -0.24 & $0.94^{+0.05}_{-0.05}$ & 0.59 & 5935 & $4.28$   & $9.58^{+1.90}_{-1.90}$ \footref{A17}      & 0.0062 & 0.1345 & $-5.19^{+0.06}_{-0.06}$ & \esp          \\
KIC 5774694  & 0.07 & $1.06^{+0.05}_{-0.06}$ & 0.64 & 5875 & $4.47$   & $1.90^{+1.90}_{-1.60}$ \footref{C14}      & 0.0139 & 0.2714 & $-4.57^{+0.02}_{-0.02}$ & \esp          \\
KIC 5774694  & 0.07 & $1.06^{+0.05}_{-0.06}$ & 0.64 & 5875 & $4.47$   & $1.90^{+1.90}_{-1.60}$ \footref{C14}      & 0.0141 & 0.2839 & $-4.55^{+0.01}_{-0.01}$ & \narval       \\
KIC 8760414  & -1.14 & $0.81^{+0.03}_{-0.02}$ & 0.52 & 5787 & $4.33$   & $11.66^{+1.61}_{-1.61}$  \footref{A17}    & 0.0069 & 0.1516 & $-5.01^{+0.04}_{-0.04}$ & \narval       \\
KIC 9955598  & 0.11 & $0.90^{+0.04}_{-0.03}$ & 0.72 & 5410 & $4.48$   & $6.29^{+1.84}_{-1.84}$ \footref{A17}      & 0.0083 & 0.1763 & $-4.93^{+0.04}_{-0.05}$& \narval       \\
KIC 7970740  & -0.49 & $0.73^{+0.03}_{-0.01}$ & 0.74 & 5290 & $4.58$   & $12.98^{+2.00}_{-2.00}$ \footref{A17}     & 0.0074 & 0.151 & $-5.05^{+0.07}_{-0.08}$ & \esp          \\
\hline
\end{tabular}
\captionof{table}{Results for stars with effective temperatures less than 6200 K as shown in Figure \ref{calcium_emission_plot}. Stars are listed in order of decreasing effective temperature. $(B-V)$ values taken from SIMBAD \citep{2000A&AS..143....9W}. Metallicity, surface gravity and effective temperature values are taken from \citet{2012MNRAS.423..122B}. The uncertainties from \citet{2012MNRAS.423..122B} are 60K for $T_{eff}$, 0.03 dex for $log(g)$ and 0.06 dex for [Fe/H]. Ages are taken from asteroseismology studies \citep{2014ApJS..210....1C,2017ApJ...835..173S}. Stars with observations from both spectrographs are recorded here for completeness.}
}
\egroup
\end{minipage}

\newpage

%Hot stars table
\begin{minipage}{\linewidth}
\centering
%\begin{small}
\bgroup
\def\arraystretch{1.4}
{
\begin{tabular}{llcccccccccl}
\hline
Star Name    & [Fe/H] & M / $M_{\odot}$ & $B-V$  & $T_{eff}$ / K & $log(g)$ & Age / Gyr & 	S Index      & \smw  & $log(R^{'}_{HK})$ & Spectrograph \\
\hline
KIC 9226926  & -0.23 & $1.39^{+0.05}_{-0.04}$ & 0.41 & 6892 & $4.14$   & $2.20^{+0.30}_{-0.30}$ \footref{C14}      & 0.0089 & 0.1876 & $-4.80^{+0.01}_{-0.01}$ & \narval       \\
KIC 3733735  & -0.04 & $1.39^{+0.04}_{-0.05}$ & 0.43 & 6715 & $4.26$   & $0.80^{+0.40}_{-0.40}$ \footref{C14}      & 0.0074 & 0.1549 & $-4.98^{+0.04}_{-0.05}$ & \esp          \\
KIC 2837475  & -0.02 & $1.43^{+0.02}_{-0.02}$ & 0.45 & 6700 & $4.16$   & $1.63^{+0.18}_{-0.18}$ \footref{A17}      & 0.0076 & 0.1584 & $-4.96^{+0.04}_{-0.04}$ & \esp          \\
KIC 7529180  & -0.02 & $1.39^{+0.04}_{-0.03}$ & 0.42 & 6700 & $4.23$   & $1.30^{+0.30}_{-0.30}$ \footref{C14}      & 0.0091 & 0.1918 & $-4.77^{+0.01}_{-0.01}$ & \narval       \\
KIC 9206432  & 0.23 & $1.38^{+0.04}_{-0.02}$ & 0.44 & 6608 & $4.23$   & $1.53^{+0.30}_{-0.30}$ \footref{A17}      & 0.0057 & 0.1248 & $-5.29^{+0.10}_{-0.14}$ & \esp          \\
KIC 11253226 & -0.08 & $1.41^{+0.02}_{-0.01}$ & 0.39 & 6605 & $4.16$   & $1.60^{+0.13}_{-0.13}$ \footref{A17}      & 0.0073 & 0.1547 & $-5.00^{+0.04}_{-0.05}$ & \esp          \\
KIC 1430163  & -0.11 & $1.34^{+0.06}_{-0.06}$ & 0.49 & 6520 & $4.22$   & $1.90^{+0.60}_{-0.50}$ \footref{C14}      & 0.0078 & 0.1675 & $-4.90^{+0.02}_{-0.02}$ & \narval       \\
KIC 10709834 & -0.08 & $1.39^{+0.04}_{-0.03}$ & 0.44 & 6508 & $4.09$   & $2.80^{+0.40}_{-0.40}$ \footref{C14}      & 0.0084 & 0.1784 & $-4.84^{+0.03}_{-0.03}$ & \narval       \\
KIC 9139163  & 0.15 & $1.40^{+0.03}_{-0.02}$ & 0.48 & 6400 & $4.18$   & $1.60^{+0.22}_{-0.22}$ \footref{A17}      & 0.0055 & 0.1226 & $-5.32^{+0.07}_{-0.09}$ & \esp          \\
KIC 9139163  & 0.15 & $1.40^{+0.03}_{-0.02}$ & 0.48 & 6400 & $4.18$   & $1.60^{+0.22}_{-0.22}$ \footref{A17}      & 0.0064 & 0.1412 & $-5.09^{+0.03}_{-0.03}$ & \narval       \\
KIC 8179536  & 0.01 & $1.16^{+0.05}_{-0.06}$ & 0.50 & 6344 & $4.27$   & $3.54^{+0.81}_{-0.81}$ \footref{A17}      & 0.0075 & 0.1617 & $-4.94^{+0.04}_{-0.04}$ & \narval       \\
KIC 10016239 & -0.05 & $1.23^{+0.05}_{-0.05}$ & 0.54 & 6340 & $4.31$   & $1.80^{+0.80}_{-0.70}$ \footref{C14}      & 0.0065 & 0.1430 & $-5.09^{+0.05}_{-0.06}$ & \narval       \\
KIC 7206837  & 0.14 & $1.30^{+0.03}_{-0.03}$ & 0.46 & 6304 & $4.17$   & $2.90^{+0.30}_{-0.30}$ \footref{A17}      & 0.0071 & 0.1540 & $-4.99^{+0.05}_{-0.06}$ & \narval       \\
KIC 1435467  & -0.01 & $1.32^{+0.03}_{-0.05}$ & 0.43 & 6264 & $4.09$   & $3.02^{+0.35}_{-0.35}$ \footref{A17}      & 0.0068 & 0.1486 & $-5.03^{+0.02}_{-0.02}$ & \narval       \\
KIC 6225718  & -0.17 & $1.16^{+0.03}_{-0.03}$ & 0.50 & 6230 & $4.32$   & $2.41^{+0.43}_{-0.43}$ \footref{A17}      & 0.0063 & 0.1397 & $-5.11^{+0.03}_{-0.03}$ & \narval       \\
\hline
\end{tabular}
\egroup
\captionof{table}{Results for stars with effective temperatures greater than 6200 K as shown in Figure \ref{calcium_emission_plot}. Stars are listed in order of decreasing effective temperature. $(B-V)$ values taken from SIMBAD \citep{2000A&AS..143....9W}. Metallicity, surface gravity and effective temperatures taken from \citet{2012MNRAS.423..122B}. The uncertainties from \citet{2012MNRAS.423..122B} are 60K for $T_{eff}$, 0.03 dex for $log(g)$ and 0.06 dex for [Fe/H]. Ages are taken from asteroseismology studies \citep{2014ApJS..210....1C,2017ApJ...835..173S}. Stars with observations from both spectrographs are recorded here for completeness.}
}

\end{minipage}
\end{landscape}

\section{List of Calibrator Stars}
\label{list_calibrator_stars}

%ESPADONS TABLE
\begin{minipage}{\linewidth}
\centering
%\begin{small}
\bgroup
\def\arraystretch{1.4}
{
\begin{tabular}{llccc}
\hline
Star Name          & Spectral Type & $B-V$  & Average \smw & \esp S Index \\
\hline
Cl Melotte 22 1613 & G0            & 0.54 & 0.3077 & $0.0119^{+0.0006}_{-0.0006}$\\
HD 124897          & K1.5III       & 1.23 & 0.1441 & $0.0075^{+0.0001}_{-0.0001}$\\
HD 144579          & G8V           & 0.73 & 0.1628 & $0.0081^{+0.0005}_{-0.0005}$\\
HD 159181          & G2Ib-IIa      & 0.98 & 0.3019 & $0.0167^{+0.0002}_{-0.0002}$\\
HD 171635          & F7Ib          & 0.62 & 0.0499 & $0.0017^{+0.00002}_{-0.00002}$\\
HD 177241          & G9IIIb        & 1.00 & 0.1057 & $0.0053^{+0.0001}_{-0.0001}$\\
HD 179958          & G2V           & 0.65 & 0.1571 & $0.0068^{+0.0002}_{-0.0002}$\\
HD 182572          & G7IV          & 0.77 & 0.1543 & $0.0071^{+0.0002}_{-0.0002}$\\
HD 185351          & G8.5IIIb      & 0.94 & 0.1946 & $0.0116^{+0.0001}_{-0.0001}$\\
HD 186408          & G1.5Vb        & 0.64 & 0.1476 & $0.0071^{+0.0001}_{-0.0001}$\\
HD 186427          & G3V           & 0.66 & 0.1450 & $0.0073^{+0.0003}_{-0.0003}$\\
HD 187691          & F8V           & 0.56 & 0.1479 & $0.0059^{+0.0004}_{-0.0004}$\\
HD 192876          & G3Ib          & 1.07 & 0.2510 & $0.0129^{+0.0001}_{-0.0001}$\\
HD 194093          & F8Ib          & 0.67 & 0.0564 & $0.0039^{+0.0005}_{-0.0005}$\\
HD 204867          & G0Ib          & 0.82 & 0.1617 & $0.0074^{+0.0001}_{-0.0001}$\\
HD 206859          & G5Ib          & 1.17 & 0.2041 & $0.0140^{+0.0002}_{-0.0002}$\\
HD 209750          & G2Ib          & 0.96 & 0.2081 & $0.0117^{+0.0001}_{-0.0001}$\\
HD 217014          & G2IV          & 0.70 & 0.1548 & $0.0068^{+0.0002}_{-0.0002}$\\
HD 26630           & G0Ib          & 0.96 & 0.3623 & $0.0148^{+0.0001}_{-0.0001}$\\
HD 31910           & G1Ib          & 0.93 & 0.2659 & $0.0124^{+0.0001}_{-0.0001}$\\
HD 3795            & K0V           & 0.70 & 0.1587 & $0.0069^{+0.0001}_{-0.0001}$\\
HD 39587           & G0V           & 0.60 & 0.3067 & $0.0173^{+0.0001}_{-0.0001}$\\
HD 7924            & K0.5V         & 0.85 & 0.2221 & $0.0102^{+0.0001}_{-0.0001}$\\
HD 89449           & F6IV-V        & 0.44 & 0.1781 & $0.0078^{+0.0001}_{-0.0001}$\\
HD 98736		   & G5			   & 0.84 & 0.1940 & $0.0096^{+0.0001}_{-0.0001}$\\
Parenago 1394	   & F8			   & 0.48 & 0.2635 & $0.0141^{+0.0003}_{-0.0003}$\\
\hline
\end{tabular}
\captionof{table}{List of calibrator stars \citep{1991ApJS...76..383D} used to calibrate the relationship between the S index calculated from \esp spectra and \smw. Spectral types and $B-V$ taken from SIMBAD \citep{2000A&AS..143....9W}.}
}
\egroup
\end{minipage}

\clearpage
\newpage

%NARVAL TABLE
\begin{minipage}{\linewidth}
\centering
%\begin{small}
\bgroup
\def\arraystretch{1.4}
{
\begin{tabular}{llccc}
\hline
Star Name         & Spectral Type & $B-V$  & Average \smw & \narval S Index \\
\hline
Cl Melotte 111 58 & F7V           & 0.45 & 0.2903        & $0.0159^{+0.0002}_{-0.0002}$ \\
Cl Melotte 111 97 & F9V           & 0.55 & 0.2978        & $0.0169^{+0.0002}_{-0.0002}$ \\
HD 103095         & K1V           & 0.75 & 0.1828        & $0.0093^{+0.0001}_{-0.0001}$ \\
HD 109358         & G0V           & 0.61 & 0.1774        & $0.0074^{+0.0001}_{-0.0001}$ \\
HD 114710         & F9.5V         & 0.59 & 0.2000        & $0.0087^{+0.0001}_{-0.0001}$ \\
HD 126660         & F7V           & 0.51 & 0.2500        & $0.0124^{+0.0001}_{-0.0001}$ \\
HD 145675         & K0V           & 0.90 & 0.1514        & $0.0074^{+0.0001}_{-0.0001}$ \\
HD 148816         & F7V           & 0.53 & 0.1621        & $0.0068^{+0.0001}_{-0.0001}$ \\
HD 150680         & G0IV          & 0.63 & 0.1354        & $0.0064^{+0.0001}_{-0.0001}$ \\
HD 154345         & G8V           & 0.76 & 0.1927        & $0.0084^{+0.0001}_{-0.0001}$ \\
HD 154417         & F8.5IV-V      & 0.58 & 0.2706        & $0.0130^{+0.0001}_{-0.0001}$ \\
HD 161797         & G5IV          & 0.75 & 0.1359        & $0.0076^{+0.0002}_{-0.0002}$ \\
HD 162003         & F5IV-V        & 0.44 & 0.1669        & $0.0075^{+0.0001}_{-0.0001}$ \\
HD 164922         & G9V           & 0.80 & 0.1598        & $0.0075^{+0.0001}_{-0.0001}$ \\
HD 16895          & F8V           & 0.51 & 0.1587        & $0.0066^{+0.0001}_{-0.0001}$ \\
HD 171635         & F7Ib          & 0.62 & 0.0499        & $0.0018^{+0.0001}_{-0.0001}$ \\
HD 176377         & G1V           & 0.59 & 0.1772        & $0.0089^{+0.0001}_{-0.0001}$ \\
HD 177830         & K0+M4V        & 1.09 & 0.1233        & $0.0059^{+0.0001}_{-0.0001}$ \\
HD 183650         & G5            & 0.71 & 0.1237        & $0.0055^{+0.0002}_{-0.0002}$ \\
HD 184499         & G0V           & 0.58 & 0.1449        & $0.0070^{+0.0001}_{-0.0001}$ \\
HD 186760         & G0V           & 0.58 & 0.1493        & $0.0060^{+0.0001}_{-0.0001}$ \\
HD 188512         & G8IV          & 0.85 & 0.1414        & $0.0060^{+0.0003}_{-0.0003}$ \\
HD 19373          & G0V           & 0.59 & 0.1517        & $0.0066^{+0.0001}_{-0.0001}$ \\
HD 194093         & F8Ib          & 0.67 & 0.0564        & $0.0018^{+0.0001}_{-0.0001}$ \\
HD 204867         & G0Ib          & 0.82 & 0.1617        & $0.0075^{+0.0001}_{-0.0001}$ \\
HD 20630          & G5V           & 0.67 & 0.3545        & $0.0182^{+0.0002}_{-0.0002}$ \\
HD 209750         & G2Ib          & 0.96 & 0.2081        & $0.0117^{+0.0001}_{-0.0001}$ \\
HD 218209         & G6V           & 0.65 & 0.1742        & $0.0080^{+0.0003}_{-0.0003}$ \\
HD 219834         & G8.5IV        & 0.79 & 0.1645        & $0.0068^{+0.0001}_{-0.0001}$ \\
HD 22072          & K0.5IV        & 0.89 & 0.1359        & $0.0060^{+0.0002}_{-0.0002}$ \\
HD 221170         & G2IV          & 1.08 & 0.1061        & $0.0050^{+0.0002}_{-0.0002}$ \\
HD 222368         & F7V           & 0.50 & 0.1520        & $0.0070^{+0.0002}_{-0.0002}$ \\
HD 22484          & F9IV-V        & 0.85 & 0.1469        & $0.0068^{+0.0001}_{-0.0001}$ \\
HD 25329          & K1V           & 0.87 & 0.1904        & $0.0083^{+0.0002}_{-0.0002}$ \\
HD 26630          & G0Ib          & 0.96 & 0.3623        & $0.0145^{+0.0001}_{-0.0001}$ \\
HD 42807          & G2V           & 0.68 & 0.3518        & $0.0166^{+0.0004}_{-0.0004}$ \\
HD 4614           & F9V+M0-V      & 0.58 & 0.1596        & $0.0073^{+0.00002}_{-0.00002}$ \\
HD 82885          & G8Va          & 0.77 & 0.3082        & $0.0152^{+0.0002}_{-0.0002}$ \\
HD 9927           & K3-III        & 1.28 & 0.1167        & $0.0061^{+0.0002}_{-0.0002}$ \\
\hline
\end{tabular}
\captionof{table}{List of calibrator stars \citep{1991ApJS...76..383D} used to calibrate the relationship between the S index calculated from \narval spectra and \smw. Spectral types and $B-V$ taken from SIMBAD \citep{2000A&AS..143....9W}.}
}
\egroup
\end{minipage}

\section{Results from multiple observations}
\label{multiple_obs_tables_appendix}

%Cool Stars <6500K
%\begin{minipage}{\linewidth}
\begin{tabular}{llllll}
\hline
Star Name    & Spectrograph & Year & S Index (mod)      & \smw & $log(R^{'}_{HK})$ \\
\hline
KIC 6106415  & \narval       & 2010 & 0.0062 & 0.1498  & $-5.03^{+0.02}_{-0.02}$\\
             & \esp          & 2014 & 0.0067 & 0.1640 & $-4.94^{+0.02}_{-0.02}$\\
KIC 8694723  & \narval       & 2010 & 0.0061 & 0.1503  & $-5.02^{+0.02}_{-0.02}$\\
             & \esp          & 2014 & 0.0068 & 0.1643 & $-4.92^{+0.02}_{-0.02}$\\
KIC 9139151  & \esp          & 2010 & 0.0058 & 0.1404  & $-5.11^{+0.04}_{-0.05}$\\
             & \esp          & 2014 & 0.0069 & 0.1681 & $-4.90^{+0.04}_{-0.04}$\\
KIC 9955598  & \narval       & 2010 & 0.0060 & 0.1482 & $-5.09^{+0.04}_{-0.05}$\\
             & \esp          & 2013 & 0.0078 & 0.1849 & $-4.90^{+0.01}_{-0.01}$\\
KIC 10454113 & \narval       & 2010 & 0.0070 & 0.1691 & $-4.90^{+0.02}_{-0.02}$\\
             & \esp          & 2014 & 0.0068 & 0.1613 & $-4.94^{+0.04}_{-0.04}$\\
             & \esp          & 2016 & 0.0073 & 0.1756 & $-4.86^{+0.01}_{-0.01}$\\
KIC 10963065 & \narval       & 2010 & 0.0060 & 0.1484 & $-5.04^{+0.02}_{-0.02}$\\
             & \esp          & 2013 & 0.0062 & 0.1524 & $-5.00^{+0.01}_{-0.01}$\\
\hline
\end{tabular}
\captionof{table}{Results of stars with multiple observations and $T_{eff} < 6500$ K. S Index (mod) indicates the S index value calculated only the H and R channels.}
%\end{minipage}

\section{Plot of sample using two channels}
\label{sample_2_channels}

\begin{minipage}{\linewidth}
	\centering
	\includegraphics[scale=0.65]{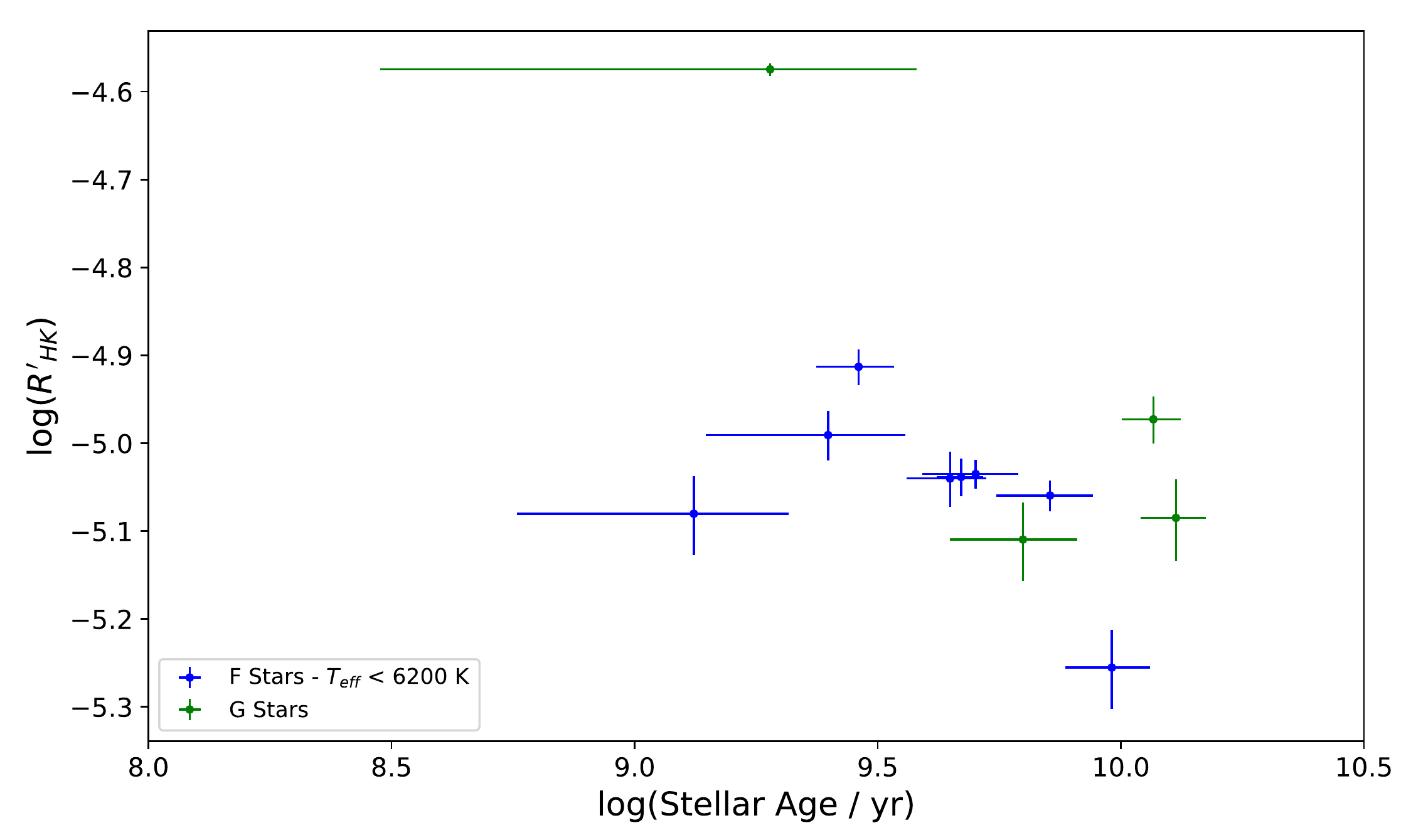}
	\captionof{figure}{Plot showing the same sample as Figure \ref{calcium_emission_plot}, but with \Rprime values calculated using only the H and R channels. This was in order to compare values with the additional observations found in the \esp archive.}
	\label{plot_sample_2channels}
\end{minipage}

%%%%%%%%%%%%%%%%%%%%%%%%%%%%%%%%%%%%%%%%%%%%%%%%%%

% Don't change these lines
\bsp	% typesetting comment
\label{lastpage}
\end{document}